\documentclass[pdflatex,sn-basic]{sn-jnl}

\usepackage{graphicx}%
\usepackage{multirow}%
\usepackage{amsmath,amssymb,amsfonts}%
\usepackage{amsthm}%
\usepackage{mathrsfs}%
\usepackage[title]{appendix}%
\usepackage{xcolor}%
\usepackage{textcomp}%
\usepackage{manyfoot}%
\usepackage{booktabs}%
\usepackage{algorithm}%
\usepackage{algorithmicx}%
\usepackage{algpseudocode}%
\usepackage{listings}%
\usepackage{lineno}%
\usepackage{pdfpages}
\theoremstyle{thmstyleone}%
\theoremstyle{thmstyletwo}%

\theoremstyle{thmstylethree}%

\begin{document}

\title[Generative AI-Downscaling of Large Ensembles Project Unprecedented Future Droughts]{Generative AI-Downscaling of Large Ensembles Project Unprecedented Future Droughts}


\author*[1,2]{\fnm{Hamish} \sur{Lewis}}\email{hamish.lewis@waikato.ac.nz}

\author[2,3,4]{\fnm{Neelesh} \sur{Rampal}}\email{neelesh.rampal@niwa.co.nz}

\author[2]{\fnm{Peter B.} \sur{Gibson}}\email{peter.gibson@niwa.co.nz}

\author[1]{\fnm{Luke J.} \sur{Harrington}}\email{luke.harrington@waikato.ac.nz}

\author[5,6]{\fnm{Chiara M.} \sur{Holgate}}\email{chiara.holgate@anu.edu.au}

\author[3,4]{\fnm{Anna} \sur{Ukkola}}\email{a.ukkola@unsw.edu.au}

\author[5,6]{\fnm{Nicola M.} \sur{Maher}}\email{nicola.maher@anu.edu.au}

\affil[1]{\orgdiv{School of Science}, \orgname{University of Waikato}, \orgaddress{\city{Hamilton}, \country{New Zealand}}}

\affil[2]{\orgname{Earth Sciences New Zealand}, \country{New Zealand}}

\affil[3]{\orgdiv{ARC Centre of Excellence for Weather of the 21st Century}, \orgname{University of New South Wales}, \city{Sydney}, \country{Australia}}

\affil[4]{\orgdiv{Climate Change Research Centre}, \orgname{University of New South Wales}, \city{Sydney}, \country{Australia}}

\affil[5]{\orgdiv{Research School of Earth Sciences}, \orgname{Australian National University}, \city{Canberra}, \country{Australia}}

\affil[6]{\orgdiv{ARC Centre of Excellence for the Weather of the 21st Century}, \orgname{Australian National University}, \city{Canberra}, \country{Australia}}

\clearpage

\maketitle


\newpage

\section*{Abstract}
Understanding how droughts may change in the future is essential for anticipating and mitigating their adverse impacts. However, robust climate projections require large amounts of high-resolution climate simulations, particularly for assessing extreme events. Here, we use a novel dataset, multiple large-ensembles of Global Climate Models (GCMs), downscaled to 12km using generative AI, to quantify the future risk of meteorological drought across New Zealand. The ensembles consists of 20 GCMs, including two single-model initial condition large ensembles. The AI is trained to emulate a physics-based regional climate model (RCM) used in dynamically downscaling, and adds a similar amount of value as the RCM across precipitation and drought metrics. Marked increases in precipitation variability are found across all ensembles, alongside highly uncertain changes in mean precipitation. Future projections show droughts will become more intense across the majority of the country, however, internal variability and model uncertainty obscure future changes in drought durations and frequency across large portions of the country. This uncertainty is understated using a smaller number of dynamically-downscaled simulations. We find evidence that extreme droughts up to twice as long as those found in smaller ensembles, could occur across the entirety of the country in the current climate, highlighting the value of long-duration downscaled simulations to sample rare events. These extremely long droughts increase in length in many locations under a high emissions SSP3-7.0 scenario giving rise to events around 30 months long in some locations.


\clearpage
\section{Introduction}\label{sec1}

Droughts are one of the most impactful natural climate hazards, causing significant excess mortality \citep{douris2021atlas}, economic losses \citep{naumann2021increased,zaveri2023droughts}, threatening water \citep{sousa2018day} and food security \cite{vogel2019effects,lesk2016influence}, and causing lasting environmental damage \citep{vicente2020review}. Anthropogenic climate change is set to increase drought risk across many regions of the globe \citep{singh2022enhanced,Falster2024,samaniego2018anthropogenic,van2023large,stevenson2022twenty}. However, for some countries such as New Zealand, future drought risks still remain uncertain \citep{Lewis2025,gibson2025downscaled} . \\

Quantifying future drought risk in New Zealand is particularly difficult due to highly uncertain summertime precipitation projections, with both model uncertainty (uncertainty arising from the use of different climate models) and internal variability \citep[natural fluctuations in the climate system,][]{deser2012uncertainty,deser2014} obscuring the sign and magnitude of change across CMIP6 Global Climate Models (GCMs) \citep{gibson2024storylines}. New Zealand's complex topography and land-sea contrasts means that GCMs often struggle to simulate both extremes, and finer scale climatic features, \citep{rummukainen2016added,gibson2024dynamical}. Thus, dynamically-downscaled climate projections are often required to produce actionable insights. Within he latest dynamically-downscaled CMIP6 projections for New Zealand \citep{gibson2024dynamical}, only across 50-60\% of the NZ land surface do five out of the six downscaled models agree on the sign of change of various meteorological drought metrics, while DJF precipitation sign agreement is approximately 30\% \citep{gibson2025downscaled}. \cite{Lewis2025} used a storylines approach to explore the two extremes of future precipitation projections on soil-moisture droughts, using the same downscaled projections as \cite{gibson2025downscaled}. They found that increasing future precipitation alleviated the increased actual evaporative drying of a warmer world, leading to a small increase in overall drought severity. Decreasing future precipitation compounded increased actual evapotranspiration, leading to the average year in the future becoming comparable to the driest years of the current climate, with the worst future years exhibiting unprecedented severity. Hydrological projections of New Zealand river flows have also found similar uncertainty in the sign of change across downscaled models across CMIP5 models \citep{mullen2018,collins2018hydrological}. Studies which examine synoptic conditions responsible for meteorological droughts have found that these conditions are projected to become more prevalent under climate change \citep{harrington2016,gibson2016}. These studies highlight that there are competing factors involved in producing future drought outcomes. In the face of considerable uncertainty \citep{gibson2024storylines,gibson2025downscaled,collins2018hydrological,Lewis2025} and possibly unprecedented future droughts \citep{Lewis2025}, there is considerable interest in better understanding the possible future outcomes of drought across New Zealand — outcomes that cannot be fully quantified using a small ensemble of downscaled GCMs.  \\

Single Model Initial-Condition Large Ensembles (SMILEs) created by running a climate model multiple times with perturbed initial conditions \citep{bengtsson2019can,maher2021large}, are valuable tools used to quantify both uncertainty in climate projections arising from internal variability \citep{deser2012uncertainty,deser2014}, and producing more rare events to robustly estimate their changes in frequency and intensity in a warming world \citep{suarez2018internal,fischer2013robust,haugen2018estimating}. However, the coarse spatial resolution of GCMs means that they struggle to simulate key aspects of New Zealand's climate \citep{rummukainen2016added,gibson2024dynamical}. This means that GCM SMILEs are not directly fit for purpose for actionable climate projections for New Zealand at the local scale. Regional Climate Model (RCM) SMILEs on the other hand would be more suitable for local scale projections, however the high computation cost leads to smaller ensemble sizes, with fewer models and scenarios selected \citep{leduc2019climex,von2019assessing,aalbers2018local}. This high computational cost further exacerbates challenges in producing robust climate projections faced by smaller countries with limited resources, such as New Zealand. \\ 

Artificial Intelligence (AI) based RCM emulators (AI-RCME), which are are orders of magnitude faster than RCMs themselves, are an emerging solution to tackle these challenges \citep{rampal2024enhancing,doury2023regional,chadwick2011artificial}. AI-RCME are deep learning models trained to learn the relationship between large-scale GCM meteorological fields (winds, temperature, humidity) to already dynamically-downscaled surface variables \citep[precipitation, temperature, etc.][]{rampal2024enhancing}. Recent studies have demonstrated that AI-based emulators are capable of capturing observed, historical climate characteristics including means, variability, and extremes, as well as climate change signals for mean precipitation \citep{doury2024suitability,rampal2025reliable} and temperature \citep{doury2023regional,bano2024transferability,rampal2025b}. The first use of SMILEs generated using AI-RCME to produce climate projections has  recently been published, showing significant internal variability in projections of temperature and precipitation extremes at fine scales across New Zealand \citep{rampal2025b}.  \\ 

The ability of AI-RCME to capture climatic means, variability, climate change signals \citep{doury2024suitability,rampal2025reliable}, and extremes \citep{rampal2025b} of precipitation, means that they also should in principle be able to simulate meteorological droughts. In this paper we utilize the AI-RCME SMILEs introduced in \cite{rampal2025b}, leveraging over 15,000 years of downscaled climate simulations to produce comprehensive meteorological drought projections for New Zealand. We then use this extremely large dataset to examine how both rare, and extremely rare droughts evolve in the future. The remainder of this paper is structured as follows: In Section 2.1 we outline the AI-RCME dataset used in  this analysis, in Section 2.2 we discuss the drought definitions used in this study, and in Section 2.3 we present the added value metrics used to evaluate the AI-RCME. In Section 3.1 we present the added value analysis of AI-RCME, and compare this added value to a traditional RCM, in Section 3.2 we show climate projections of meteorological drought for New Zealand using the entirety of the AI-RCME output, and in Section 3.3 we investigate the emergence of extreme future droughts. In Section 4 the results and their implications are discussed, and we present our conclusions in Section 5.

\section{Methods}\label{sec2}

\subsection{Data}

 \subsubsection{Regional Climate Model Training Data}
\label{RCM_data}

Six models from the CMIP6 ensemble \citep{Eyring2016} have recently been dynamically downscaled from their native resolutions to a resolution of 12 km over New Zealand \citep{gibson2024dynamical}. Downscaling was performed with the Conformal Cubic Atmospheric Model \citep[CCAM,][]{mcgregor2008updated} which implements a variable resolution conformal-cubic grid enhancing the resolution over an area of interest (New Zealand and its surrounding ocean), accompanied by a relatively high resolution (12-35km) over the wider South Pacific. \\ 

The six GCMs within this downscaled ensemble were chosen based on their performance across four main categories as outlined in \cite{gibson2024dynamical}: (1) GCMs similarity to observations (VCSN, \cite{tait2006,tait2012}) of annual and seasonal means of surface variables: precipitation, surface air temperature, and mean sea level pressure (MSLP). (2) Correlation of the Southern Oscillation Index (SOI) to observations of precipitation, surface air temperature, and mean sea level pressure (MSLP). (3) Annual cycle in MSLP differences used to calculate regional circulation indices Z1, Z2, M1 \citep{trenberth1976fluctuations}, and the SOI. (4) Position of the southern hemispheric jet in both summer and winter. \\ 

Three GCMs (ACCESS-CM2, EC-Earth3, NorESM2-MM) were downscaled through spectral nudging to the host GCM's atmospheric fields, sea surface temperatures (SSTs), and sea ice concentrations (SICs). Three other GCMs (AWI-CM-1-1-MR, CNRM-CM6-1, GFDL-ESM4) were downscaled in an ``AMIP'' configuration, driving CCAM with bias-corrected host model SSTs and SICs. In this analysis, we focus on comparisons between the outputs of the spectrally nudged RCM runs, and the same AI downscaled GCMs (ACCESS-CM2, EC-Earth3, NorESM2-MM), as both are forced with the GCM large scale atmospheric fields.

\subsubsection{Generative AI Downscaled Data}

This section outlines the AI downscaled precipitation dataset first presented in \cite{rampal2025b}. The AI model used to produce the data used in this study is a deep learning-based emulator adapted by \cite{rampal2025b} from previous studies \citep{rampal2024,rampal2025reliable}. The model architecture is a residual generative adversarial network (GAN), which has two components. Firstly, a convolutional neural network (CNN), namely a U-Net architecture \citep{ronneberger2015u}, is trained
to emulate a specific variable, and captures the predictable, large-scale component of said variable driven by regional circulation. Secondly, a GAN is trained on the differences between the deterministic "smooth" prediction of the CNN and the RCM output. This residual approach improves the ability of the GAN relative to the CNN, capturing extremes and fine-scale precipitation structures, but also extrapolate to warmer climates \citep{rampal2024enhancing,rampal2025reliable,rampal2025b}. The AI model separately downscales daily precipitation as well as daily maximum temperatures over the New Zealand region (165°E–184°W, 33°S–51°S). The emulator was trained using predictor and target variables output from the CCAM regional climate model which is described in the previous section. Predictor fields of horizontal winds, temperature and specific humidity (u,v,t,q) are coarsened to 1.5°, and taken at two pressure levels 850, and 500 hPa. This coarsening of the downscaled predictor fields is known as the ``perfect framework'' \citep{doury2023regional,rampal2024}. The perfect framework is often preferable to training with GCM large-scale fields directly (the ``imperfect framework''), which is often more challenging as the RCM's mean state can deviate from that of the GCM \citep{bano2024transferability,bartok2017projected,boe2023simple,doury2024suitability}. Predictor fields are normalized using the spatio-temporal mean and standard deviation for the entirety of the training data and is consistent with previous approaches \citep{bailie2024quantile,rampal2024enhancing,rampal2025reliable}. The emulator is trained using 140 years (1960–2100) of CCAM simulations forced by ACCESS-CM2 GCM. ACCESS-CM2 was chosen as the training model as it is the CCAM downscaled model with the highest climate sensitivity, allowing the AI model to sample the largest range of future conditions, allowing it to extrapolate better. \\

This AI model was then used to downscale 20 GCMs which cover both a historical period (1960-2014) and SSP3-7.0 future period (2015-2099), with the choice of GCMs based on the availability of daily large-scale predictor data. Within these 20 GCMs SMILEs of CanESM5 (n=19) and ACCESS-ESM1-5 (n=40) were also produced. For a summary of the models used in the downscaled ensemble please see \cite{rampal2025b}. It takes approximately three minutes to downscale one historical simulation, and four minutes to downscale an SSP3-7.0 simulation on an A100 GPU, totaling approximately 6 hours to generate the entire dataset using 4 A100 GPUs using this AI model. No bias correction applied to the output to maintain consistency with the RCM output. Please see \cite{rampal2025b} for a complete description of the dataset. \\

\subsection{Drought Metrics}
\label{Drought_Metrics}

With the given set of outputs produced by the AI-RCME we are able to investigate meteorological droughts (precipitation deficits) in unprecedented detail. Metrics of meteorological drought are generally considered to have the best agreement among GCMs at global scales compared to runoff and soil moisture droughts \citep{ukkola2018evaluating}. We produce meteorological drought projections using the metrics (intensity, duration, and frequency) described in \cite{ukkola2020}, as we have good confidence that these are meaningfully represented by the GCMs which are then downscaled, and have also been used previously for New Zealand \citep{gibson2025downscaled} . \\

\cite{ukkola2020} defines a drought event as a month, or a number of consecutive months, where the 3-month running sum of precipitation is lower than the 15th percentile, defined in each month of the year respectively. These 15th percentiles are computed for each grid cell across the 1960-2014 period within each downscaled historical simulation. Drought intensity is defined at each grid cell as the average deficit between the 3-monthly running sum and the 15th percentile threshold across all drought events in a set period. Drought frequency the number of drought events per year within a set period. Drought duration is defined as the the average duration of a drought event within a set period. A 3-month running sum below the 15th percentile would have a drought duration of 1 month, and two such months in a row would have a duration of 2 months, reflecting the number of months spent in drought conditions rather than in a precipitation deficit. In the case of this study, this set period is 1985-2014 in the historical simulations, and 2070-2099 in SSP3-7.0 simulations.   \\

\subsection{Added Value Metrics}

Added value in a regional climate modeling context can be defined as the reduction in bias of an RCM output relative to the host GCM. Here, we compare the added value of AI downscaling relative to that of the RCM across annual and seasonal (DJF, JJA) precipitation totals, as well as the meteorological drought metrics introduced in section \ref{Drought_Metrics}. \\ 

We quantify added value using the approach and statistical measures laid out in \cite{gibson2024dynamical}, with the range of statistical measures assessed accounting for different aspects of model performance. Root-mean-square error (RMSE), and mean-absolute error (MAE) penalize the magnitude of errors. Mean-absolute-percentage error (MAPE) penalizes based on the percentage error, which is useful for characterizing error in regions of lower absolute values. Pattern correlation (Pcorr) accounts for differences in spatial distributions of climatologies. Finally, Land\% is defined as the fraction of all grid cells where the RCM/AI provides added value based on MAE \citep{di2020realised}. Land\% is scaled from -50\% to 50\%; a Land\% of 50 would indicate that the RCM/AI adds value across every grid cell. \\ 

The reference dataset used to assess added value was the Virtual Climate Station Network (VCSN) \citep{tait2006,tait2012} gridded station product. This 5 km resolution daily product is constructed using a a second-order trivariate thin-plate smoothing spline applied across an extensive network climate stations (approximately 1200) across New Zealand. All datasets were conservatively regridded to the 12 km CCAM grid over New Zealand for evaluation. For the purpose of model evaluation we assess all metrics over the 1985-2014 period.

\section{Results}\label{sec11}

\subsection{Evaluation of AI-RCME Simulation of Drought}

\begin{figure}
    \centering
    \makebox[\textwidth][c]{\includegraphics[width=1.2\linewidth]{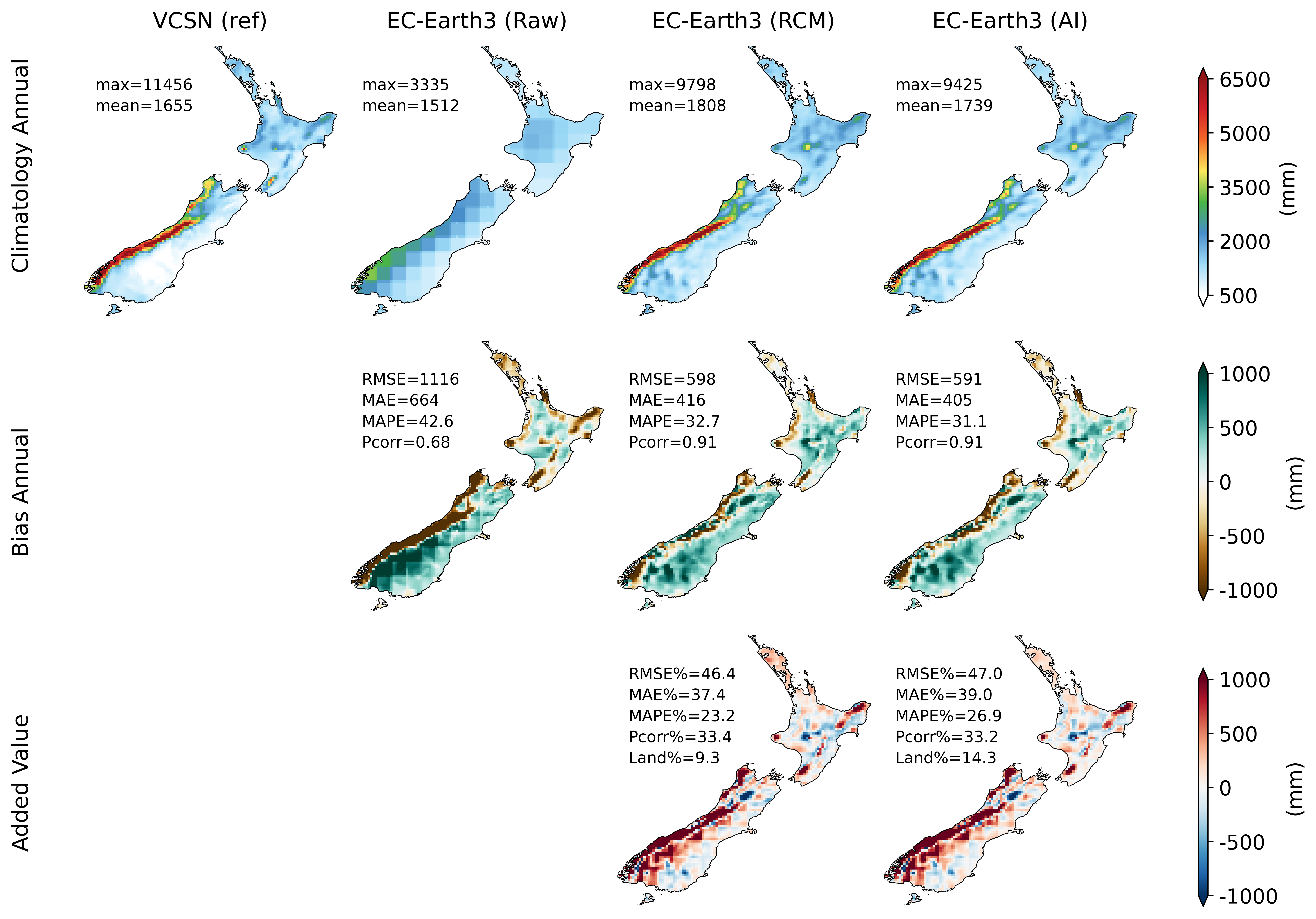}}
    \caption{Added value for the annual mean precipitation climatology (1985-2014) through downscaling the EC-Earth3 GCM with the CCAM RCM and AI-RCME. The top row shows the climatology of each dataset. The middle row shows the absolute biases between each dataset and the VCSN reference dataset, with the error metrics of each dataset shown. The bottom row shows the added value of RCM (bottom left) and AI (bottom right) downscaling, where positive values (red) indicate a reduction in absolute biases. Improvements in error metrics as percentages averaged across the land surface are also shown.}
    \label{EC-added-value}
\end{figure}

AI-RCME need to be carefully evaluated to ensure that they can capture both historical climate characteristics, and future climate change signals in their outputs \citep{doury2024suitability,rampal2025reliable}. Here, we evaluate the ability of the AI-RCME presented in \cite{rampal2025b} to simulate annual and seasonal precipitation, as well as meteorological drought metrics, over the historical period of 1985-2014. We then examine the simulated climate change signal of the meteorological drought metrics under a high emissions SSP3-7.0 scenario: SSP3-7.0. We compare the downscaling ability of the AI-RCME to that of CCAM across the three spectrally nudged GCMs: ACCESS-CM2, EC-Earth3, and NorESM2. \\

Figure \ref{EC-added-value} presents the added value of downscaling for the annual precipitation climatology of the EC-Earth3 GCM, using the VCSN product as a reference (similar figures for ACCESS-CM2 and NorESM2-MM can be found in the supplementary material which corroborate similar results). As expected the raw GCM substantially underestimates orographic precipitation across high elevation regions, particularly the windward side of the Southern Alps to the prevailing westerly. The raw GCM also has a pervasive wet bias on the leeward side of the Southern Alps, again due to the lack of orography in the GCM \citep{gibson2024dynamical,Renwick1998}. CCAM significantly reduces these orographic biases, as well as biases overall; RMSE is reduced by 46\% across the land surface, MAE by 37\%, and MAPE by 23\%. The spatial pattern of precipitation is also better represented by downscaling, with improvements in Pcorr of 33\%. Most importantly, the AI-RCME inherits all of these improvements, adding a similar amount of value as CCAM over the host GCM, reducing RMSE by 47\%, MAE by 39\%, MAPE by 26\% and improving Pcorr by 33\%. The fact that the AI-RCME adds a similar amount of value in similar places as CCAM suggests that it is reducing biases because it is correctly emulating CCAM, rather than by random chance. This is also the case for inheriting some of CCAMs biases, such as the wet bias over low elevations \citep{gibson2024dynamical}. Equivalent figures for the other nudged GCMs are presented in the supplementary material.  \\ 

\begin{figure}
    \centering
    \makebox[\textwidth][c]{\includegraphics[width=1.2\linewidth]{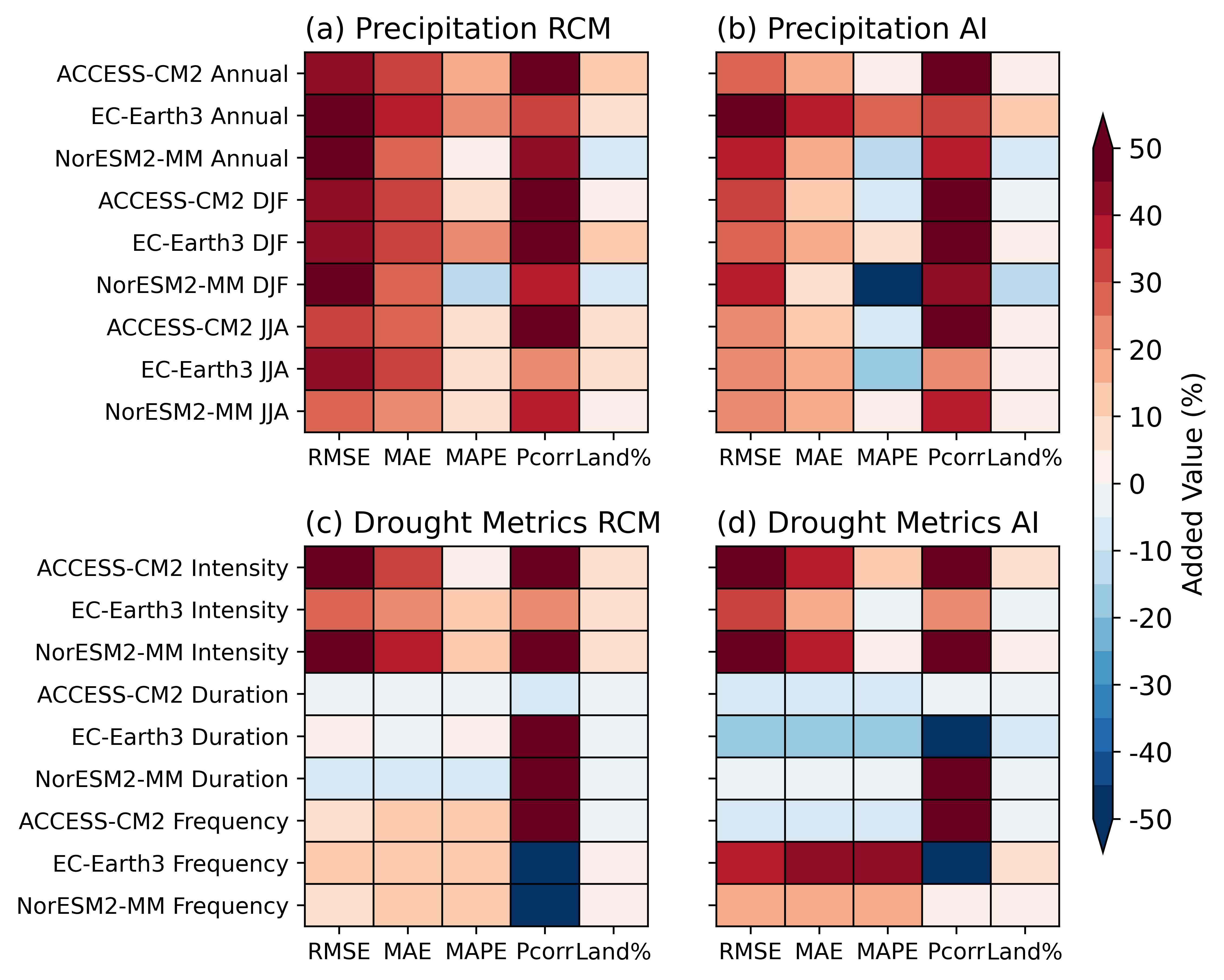}}
    \caption{Added value of downscaling  using the CCAM RCM (a,c), and AI-RCME (b,d). (a) and (b) show the added value of downscaling across annual, as well as DJF, and JJA, precipitation climatologies. (c) and (d) show the added value of downscaling across the various drought metrics discussed in section \ref{Drought_Metrics}. }
    \label{added_value_summary}
\end{figure}

The added value of CCAM and the AI-RCME across annual, and seasonal precipitation is summarized in Figure \ref{added_value_summary} (a) and (b). We see that across the nudged GCMs that the AI-RCME produces the most added value across similar metrics to CCAM, these being RMSE, MAE, and Pcorr. The AI-RCME has greater difficulty producing improvements in MAPE, possibly due to an additional wet bias over low elevations inherited from the GCMs themselves. This may be due to the AI-RCME using specific humidity (q) as a predictor where CCAM does not, and thus building relationships between q and precipitation that are not present in CCAMs downscaling. The AI-RCME produces high resolution precipitation fields that resemble CCAM, but can have similar underlying features to the host GCM, for example an additional wet bias over low elevations across the southern part of the country in ACCESS-CM2 is inherited by the AI-RCME, but is less pronounced in CCAM (see Figure S1 of the supplementary material).  Additional evidence that the AI-RCME is a high resolution intermediate between CCAM and the host GCM can be found when examining climate change signals of precipitation, where spatial changes of the AI-RCME are similar to both the host GCM and CCAM, with the host GCM and CCAM having the least similarity to each other of the three comparisons (see Figure S20 and S21 of \cite{rampal2025b}). Overall, the AI-RCME adds similar value to CCAM for annual and seasonal precipitation, albeit with slightly reduced performance, as would be expected from an emulator-based approach that cannot reproduce every aspect of the full physics-based dynamical downscaling procedure. \\ 

\begin{figure}
    \centering
    \makebox[\textwidth][c]{\includegraphics[width=0.8\linewidth]{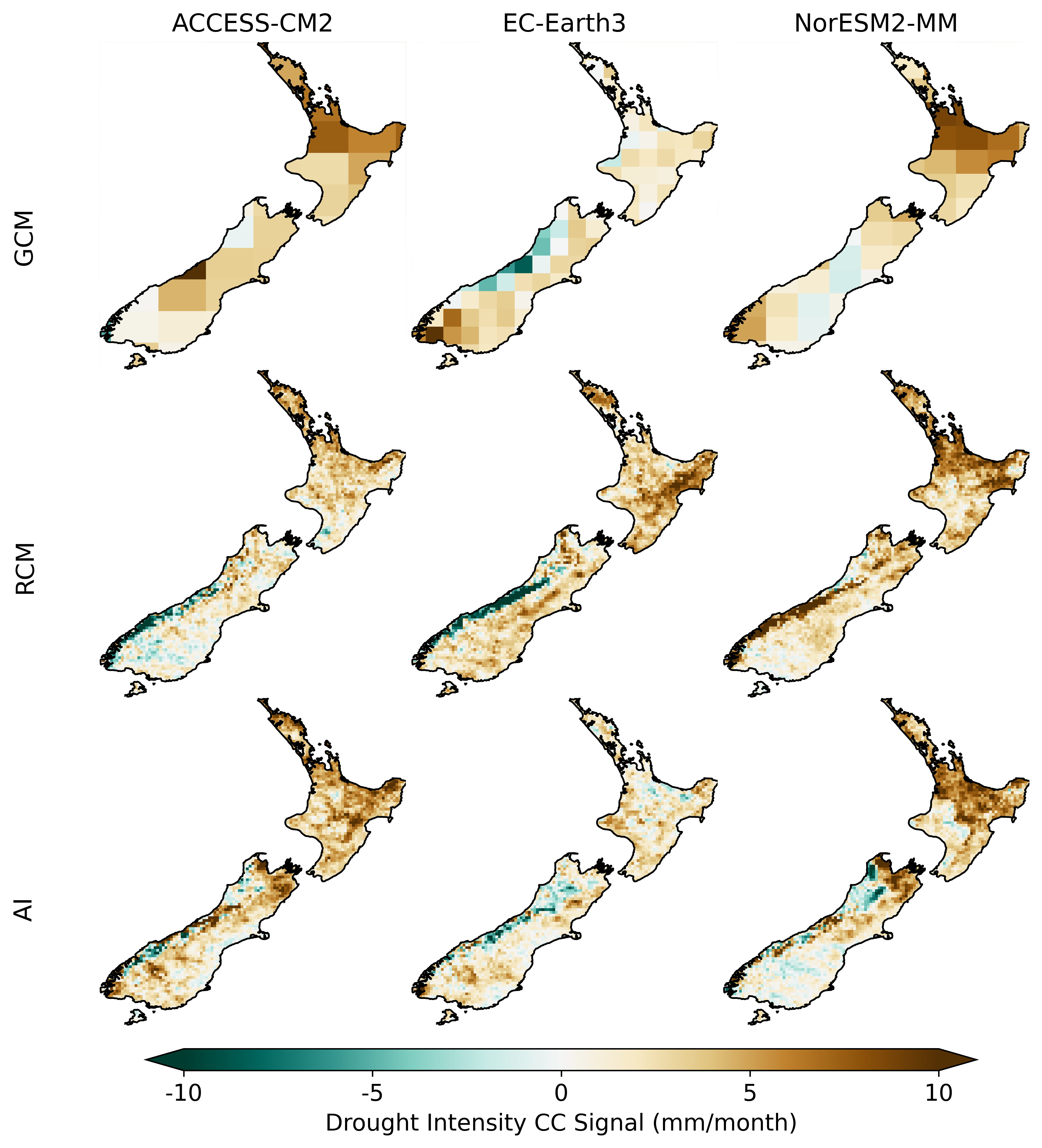}}
    \caption{Climate change signal of mean meteorological drought intensity between 2070-2099 in SSP3-7.0, relative to 1985-2014 in historical simulations. The rows show the climate change signal within the GCM, dynamically downscaled RCM output, and AI-RCME downscaled output respectively, while the columns show the host GCMs: ACCESS-CM2, EC-Earth3, and NorESM2-MM.}
    \label{cc-signals}
\end{figure}

We show the added value of both CCAM and the AI-RCME across the drought metrics outlined in Section \ref{Drought_Metrics} in Figure \ref{added_value_summary} (c) and (d). For drought intensity both CCAM and AI-RCME produce significant reductions in RMSE, MAE, and improvements in Pcorr, while having difficulty making improvements in MAPE, as shown earlier for precipitation. At first glance CCAM and AI-RCME have apparent difficulty simulating drought duration and frequency. However, this is not solely due to deficiencies in the downscaling procedure, simply, these fields are very spatially noisy and are heavily influenced by internal variability (see the internal variability present among the members of ACCESS-ESM1-5 in Figure S14 of the supplementary material for example), and thus comparisons between a models historical simulation and observations can show significant disagreement.    \\

\cite{rampal2025b} demonstrates that the AI-RCME used in this study correctly captures CCAMs climate change signal of precipitation (Figure S20 and S21). Figure \ref{cc-signals} shows the downscaled climate change signal of drought intensity between 2070-2099 in SSP3-7.0 and 1985-2014 in the historical period across the three nudged GCMs. We see that the AI-RCME broadly captures the spatial pattern of change seen in the dynamically downscaled RCM across all three GCMs. Some differences in intensity are evident in ACCESS-CM2, where the GCM, and thus the AI-RCME show significantly more intense droughts across the North Island. Similarly for EC-Earth3, where the GCM and thus the AI-RCME produce a more moderate climate change signal, the dynamically downscaled RCM is significantly drier. Equivalent figures for drought duration and frequency are presented in the supplementary material as Figures S3 and S4. Climate change signals of drought duration and frequency follow a similar relationship to those found for drought intensity, the AI-RCME broadly capturing the pattern of change seen in the RCM, with some similarities to the host GCM.    \\

Overall, downscaling using the AI-RCME adds substantial value across annual and seasonal precipitation, as well as drought intensity (compared to the host GCM), with a similar order of magnitude to CCAM, though with some expected limitations due to its emulation-based approach. Additionally, the AI-RCME can successfully capture the climate change signals of drought produced by CCAM. In summary, this high level of performance gives us confidence that the AI-RCME can broadly reproduce climatological precipitation and drought statistics similarly to CCAM, thus making its outputs suitable to produce high-resolution climate projections of drought across New-Zealand.  

\subsection{Drought Projections using AI-RCME-SMILEs}
\label{projections}

Significant uncertainty exists around the sign change of future precipitation and drought metrics at regional scales across New Zealand, in particular, precipitation across DJF and MAM, and drought durations and frequencies \citep{gibson2025downscaled,gibson2024dynamical}. One important limitation of previous regional climate projections for New Zealand is that due to computational constraints, only a single member of six GCMs was able to be downscaled. This leaves internal variability unaccounted for, while likely under-sampling model uncertainty. Both internal variability, and model uncertainty have been shown to play a large role in differing DJF precipitation projections for New Zealand within GCM climate projections \citep{gibson2024storylines}, as well as being the largest sources of uncertainty within SMILEs used for drought projections at a global scale \citep{ji2024uncertainty}. Here, we make use of the entire downscaled dataset presented in \cite{rampal2025b}, including 20 GCMs, the 19 member CanESM5 and 40 member ACCESS-ESM1-5 initial condition ensembles, to better understand seasonal precipitation projections, and meteorological drought changes at regional scales across New Zealand. \\

\begin{figure}[h!]
    \centering
    \makebox[\textwidth][c]{\includegraphics[width=1.2\linewidth]{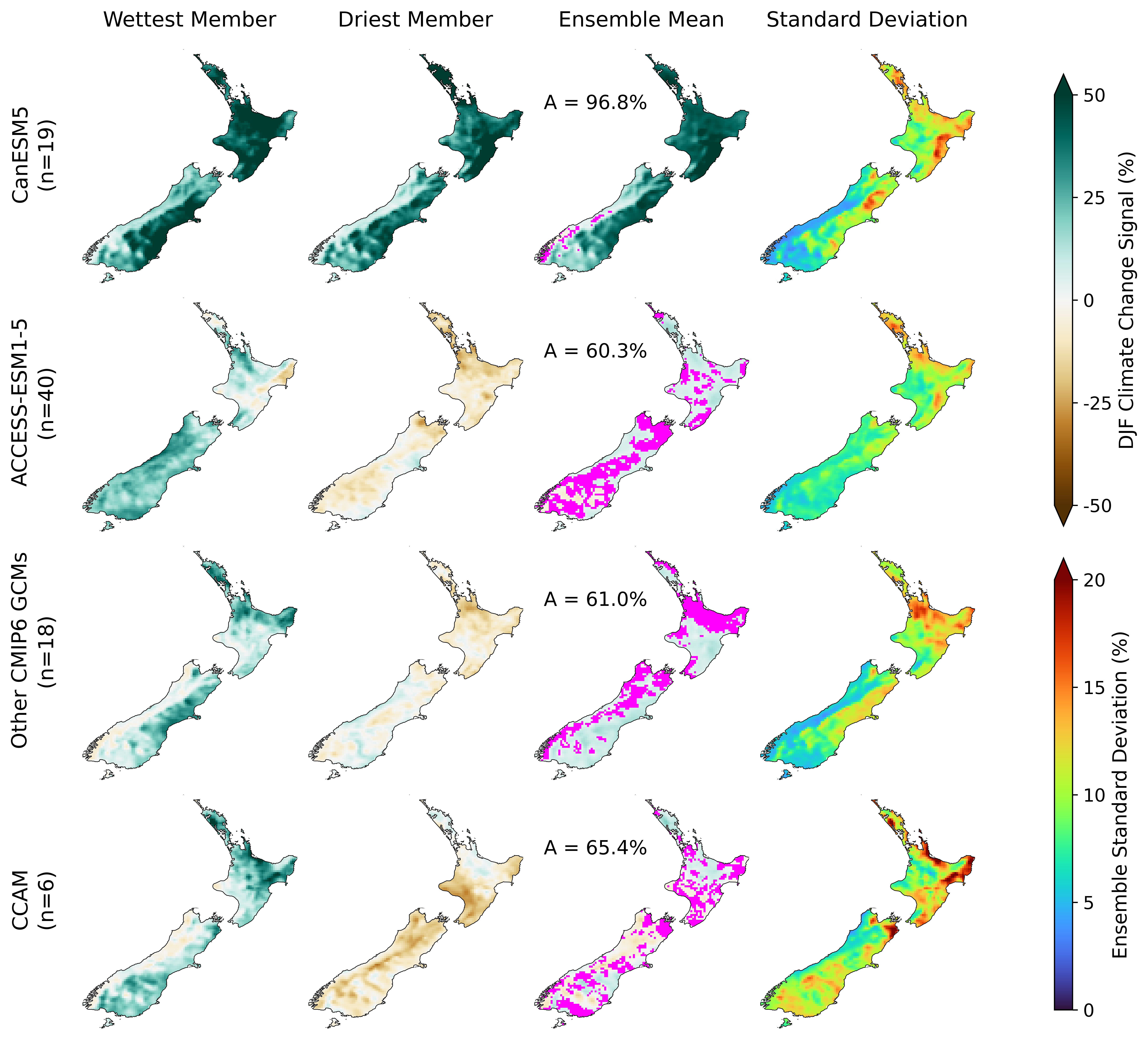}}
    \caption{DJF precipitation climate change signal between in 2070-2099 in SSP3-7.0 compared to 1985-2014 in historical simulations. The columns show the wettest member, driest member, ensemble mean, and inter-member standard deviation of each ensemble: CanESM5 (AI-downscaled), ACCESS-ESM1-5 (AI-downscaled), all other CMIP6 GCMs (AI-downscaled), and CCAM (dynamically-downscaled), which constitute the rows. Pink coloring of the ensemble means denotes grid-points where less than 66\% of ensemble members agree on the sign of change. The percentage of the land surface over which the ensemble agrees on the sign of change (A) is presented alongside the ensemble means. }
    \label{DJF_Precip}
\end{figure}

Figure \ref{DJF_Precip} presents DJF precipitation projections under an SSP3-7.0 scenario across both initial condition ensembles, the remaining 18 AI downscaled CMIP6 GCMs, as well as the GCMs dynamically downscaled with CCAM for comparison. The CanESM5 ensemble stands out due to its consistently wet climate change signal over New Zealand, with CanESM5's driest initial condition member being wetter than the wettest member of all other ensembles. Although the sign of change is in good agreement between initial condition members across the CanESM5 ensemble, internal variability obscures the exact spatial pattern of change, with ensemble standard deviation across the North Island for CanESM5 being a similar magnitude to the other ensembles, which have members whose climate change signals span zero. This large amount of variability across the initial condition members is similar to that seen in the coarse resolution GCM output shown in \cite{gibson2024storylines}. The sign change is less certain across the ACCESS-ESM1-5 ensemble with a sign agreement across 65.7\% of the land surface of a mostly wetting climate change signal. This mostly wetting signal is also seen across the other CMIP6 GCMS however the regions of sign agreement (A=61\%) are completely different to those in the ACCESS-ESM1-5 ensemble, with greater uncertainty on the west coast of the South Island and Northern part of the North Island across the other CMIP6 GCMs, and greater uncertainty in the cental and south of the South island in ACCESS-ESM1-5. These two ensembles are different again to the GCMs dynamically downscaled with CCAM, which show a drying signal across the west and south of the South Island, and a wetting signal elsewhere, ensemble agreement is less coherent and contiguous across all regions of North and South Islands (A=65.4\%). The lack of agreement on the sign change of precipitation at regional scales in DJF due to internal variability across CanESM5 and ACCESS-ESM1-5 ensembles, as well as model uncertainty across the other CMIP6 GCMs, is consistent with the findings of \cite{gibson2024storylines} across the wider New Zealand domain within GCMs.  \\

Although this new dataset cannot give additional clarity on the sign change of DJF precipitation, it does provide additional content to the existing dynamically downscaled projections presented in \cite{gibson2025downscaled}. The differing regions of ensemble agreement of sign change between ACCESS-ESM1-5, other CMIP6 GCMs, and dynamically downscaled ensembles, span the majority of the country. This suggests that the existing downscaled projections may be overconfident in the sign change of precipitation across some regions due to the undersampling of model and initial condition uncertainty. Some of the models which constitute the other 18 CMIP6 GCMs are known to poorly simulate the historical climate 
around New Zealand \citep{Gibson2023}, while the six dynamically downscaled GCMs were specifically chosen due to their performance across New Zealand \citep{gibson2024dynamical}. Thus, it is difficult to say that the dynamical downscaled GCMs substantially under-sample the ``true'' model uncertainty which is present across the other CMIP6 GCMs when such systematic biases in historical climate in some models exist. However, variability across the CanESM5 and ACCESS-ESM1-5 ensembles demonstrates that internal variability produces significant precipitation differences between ensemble members. This is difficult to quantify in the dynamically downscaled ensemble, as it is difficult to isolate its role alongside model uncertainty. Although local trends of DJF precipitation remain obscured, there is generally greater confidence across ensembles in other seasons: A=77-94\% for JJA, 75-90\% for SON, 53-85\% for MAM, and 76-100\% annually (see Figures S5-S7 in the supplementary material, and Figure \ref{drought_outcomes_nationwide}). 
 \\

\begin{figure}
    \centering
    \makebox[\textwidth][c]{\includegraphics[width=1.2\textwidth]{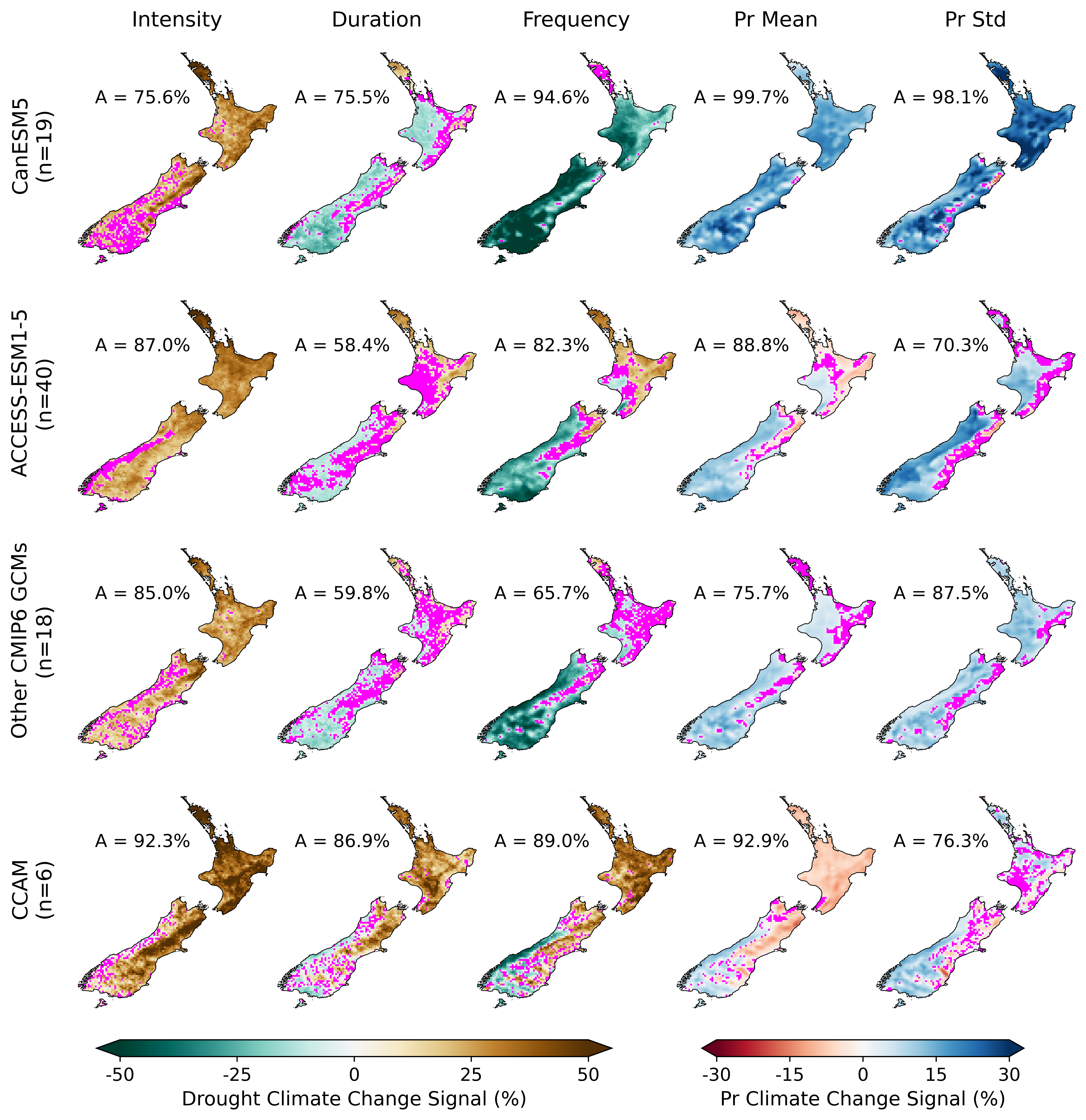}} 
    \caption{Climate change signal of various metrics between in 2070-2099 in SSP3-7.0 compared to 1985-2014 in historical simulations. The columns show the climate change signal of drought intensity, duration, frequency, as well as mean precipitation (Pr Mean), and month-to-month precipitation variability (Pr Std) throughout the whole year. The rows show the various ensembles used in this analysis: CanESM5 (AI-downscaled), ACCESS-ESM1-5 (AI-downscaled), all other CMIP6 GCMs (AI-downscaled), and CCAM (dynamically-downscaled). Pink coloring denotes grid-points where less than 66\% of ensemble members agree on the sign of change. The percentage of the land surface over which the ensemble agrees on the sign of change (A) is presented alongside. The green-brown colorbar corresponds to drought metrics, while the red-blue colorbar corresponds to precipitation metrics.}
    \label{drought_outcomes_nationwide}
\end{figure}

While mean precipitation can to some degree explain future drought outcomes across many regions of the globe, precipitation variability (defined here as the standard deviation of monthly mean precipitation totals) is also recognized as a significant driver of meteorological droughts \citep{trenberth2014global,ukkola2020}. In Figure \ref{drought_outcomes_nationwide} we present future changes in drought metrics (introduced in Section \ref{Drought_Metrics}) alongside the changes in annual mean precipitation, and precipitation variability. Drought intensity increases substantially in the future within all ensembles across the entirety of the North Island, and over large portions of the eastern South Island. While future changes in drought intensity are generally robust, their drivers between regions and ensembles can still be highly varied. Both CanESM5 and the other CMIP6 GCMs have robust increases in mean precipitation across the majority of the country, however, similar increases in precipitation variability produce more intense future droughts. The ACCESS-ESM1-5 ensemble has increases in drought intensity driven by decreases in mean precipitation over regions where changes in precipitation variability are mostly uncertain. Otherwise, increases in drought intensity are driven by increases in precipitation variability over the remainder of the country. Within the dynamically downscaled models the largest and most robust increases in drought intensity occur where large decreases in mean precipitation occur. This very strong drying signal is driven by the three simulations run in the AMIP configuration which amplify drying during spring (see Figure S26 of the supplementary material, or similarly S15 of \cite{gibson2025downscaled}).\\

While changes in drought intensity can be driven by changes in both mean precipitation, and precipitation variability, changes in drought duration and frequency are more readily driven by changes in mean precipitation \citep{ukkola2020}. This is reflected in the increases in both drought frequency and duration in the ACCESS-ESM1-5 ensemble, and in the dynamically downscaled models where mean precipitation is decreasing. The influence of substantial increases in precipitation variability can be seen to increase drought durations around the coastal areas of the North Island in the CanESM5 ensemble. Otherwise, drought durations decrease in length in regions with smaller increases in variability and larger increases in mean precipitation. \\

There is significant disagreement in the projected sign change of drought duration within the ACCESS-ESM1-5 ensemble, and drought duration and frequency within other CMIP6 GCMs. We examine the individual members of these two ensembles across the three drought metrics, mean precipitation, and precipitation variability in Figures S13-S22 of the supplementary material. There is a large amount of ensemble variance in drought durations in the ACCESS-ESM1-5 ensemble across regions that have reasonable agreement in the sign change of precipitation. This is likely due to the magnitude of the change being quite small, which does not illicit a robust change in drought durations given the relatively low signal to noise ratio. We find less robust changes in mean precipitation across the other CMIP6 GCMs, where increases in mean precipitation projected by some GCMs being relatively weak. This in turn leads to notable inter-model variability in drought durations and frequency across the North Island. \\

There is much greater agreement in projections of drought intensity across the four ensembles examined here than DJF precipitation. Although, these increases in intensity are the result of different drivers in each ensemble, all AI-downscaled ensembles corroborate the results derived from the dynamically downscaled projections, albeit with slightly less agreement across some regions due to the effects of internal variability (CanESM5, ACCESS-ESM1-5), and increased model uncertainty (other CMIP6 GCMs), which the dynamically downscaled ensemble does not capture. There is disagreement in the changes of the drought drivers themselves, primarily mean precipitation across the North Island, with CanESM5 and the other GCMs having a wetting signal, while ACCESS-ESM1-5 is dry around the northern and eastern portions of the North Island. The AI-downscaled ensembles project robust increases in precipitation across the majority of the South Island, in disagreement with the results of the dynamically downscaled ensemble. Again, these ensemble differences suggest
that the existing dynamically downscaled projections may be too confident in the sign of change of mean precipitation in certain regions, under-sampling model uncertainty and internal variability, with the caveat in the case of model uncertainty that the dynamically downscaled models were chosen due to their historical performance over New Zealand. Changes in drought duration and frequency spatially reflect those in mean precipitation, albeit with the additional influence of increasing precipitation variability greatly obscuring changes. The most notable example of this is drought durations in CanESM5, where very robust increases in both mean precipitation and precipitation variability occur across the whole country, but the sign of change of drought durations only certain across 65\% of the country.  \\

In summary, sign changes of DJF precipitation still remain obscured, with different ensembles uncertain of the sign of change across different areas of the counrty. One interpretation of this is that the dynamically downscaled ensemble may not capture a large enough range of possible DJF precipitation changes that arise from model uncertainty, and/or internal variability. Changes in annual mean precipitation are also uncertain, but with different ensembles showing opposite signs of change across the North Island and upper South Island. Precipitation variability robustly increases across most regions of all ensembles. The majority of the country is projected to have increasingly intense droughts by the end of the century driven in some regions by decreasing mean precipitation, and nationwide by increases in precipitation variability. Increases and decreases in drought durations and frequencies generally follow the spatial pattern of mean precipitation, however the ensemble agreement (A) derived from the AI-downscaled ensembles is far lower than that of the dynamically downscaled models.

\begin{figure}
    \centering
    \makebox[\textwidth][c]{\includegraphics[width=1.2\textwidth]{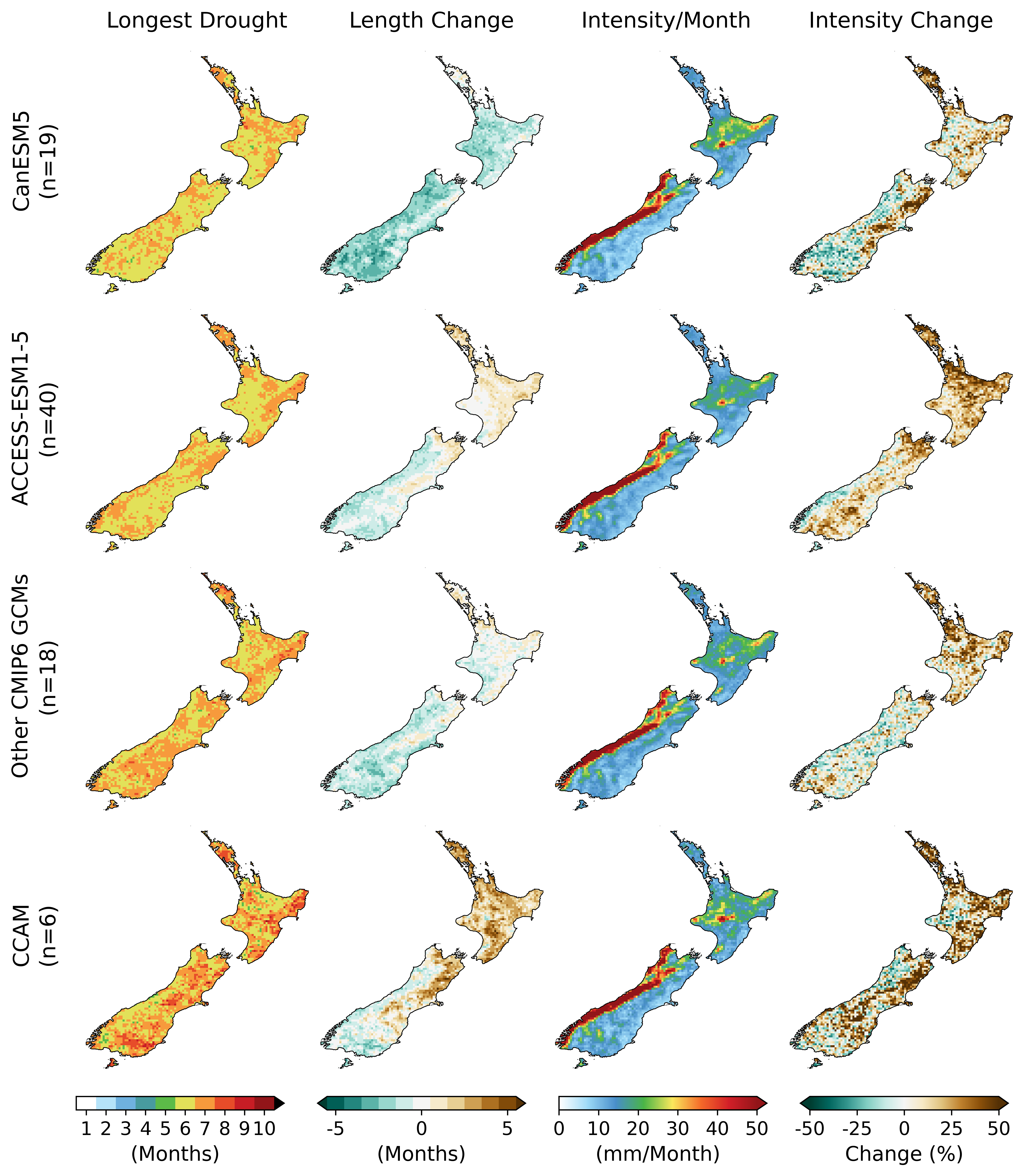}} 
    \caption{First column: Median longest drought across all members at each grid-point within an extended historical period (1965-2014). Second Column: The change in median longest drought across members in an extended future period (2050-2099) under an SSP3-7.0 scenario. Third Column: Median intensity per month of the longest droughts across members. Fourth Column: Change in the median intensity per month of the longest drought across members under an SSP3-7.0 scenario. }
    \label{median_drought}
\end{figure}

\subsection{Extreme Droughts in Current and Future Climates}

One of the key strengths of large ensembles, is that the greater number of model years available allow for more robust estimates of the intensity and frequency of extreme events \citep{suarez2018internal,fischer2013robust,haugen2018estimating}. Here, we again leverage the entire AI-downscaled dataset presented in \cite{rampal2025b} (over 15000 years) to better estimate the severity of extreme droughts in the current climate, as well as how these extreme droughts may change in the future under a high emission SSP3-7.0 scenario. We choose to examine the longest droughts across these ensembles in two different frameworks over a greater number of years: 1965-2014 in the historical period, and 2050-2099 within SSP3-7.0 simulations, to further increase our sample size of events. Firstly, we find the longest drought present within each member of each ensemble at each grid-point in these periods. We then examine changes in the median longest drought among members to provide a meaningful comparison between ensembles with a different number of members, as well as providing an estimate of severe events that would be likely in a 50 year period. Secondly, we examine the longest drought across the members of each ensemble at each grid-point, to gain an understanding of the most severe events possible in the current climate, and how these extremely long droughts will change in the future under a high emission SSP3-7.0 scenario. \\ 

Columns one and three of Figure \ref{median_drought} show the length and intensity of the median longest drought across members at each grid-point in an extended historical period (1965-2014). Across ensembles the median longest drought is consistently 6-7 months in length with a deficit of approximately 15mm a month below the 15th percentile across low-lying regions ($<$1000m) where pastoral agriculture is most prevalent in New Zealand. This 15mm per month deficit corresponds to approximately 40\% less precipitation than average over this 6-7 month span across these low-lying regions. The duration and intensity change of the median longest drought in an extended future period (2050-2099) are shown in columns 2 and 4 of Figure \ref{median_drought}.
Changes in drought intensity are robust across ensembles, with a broadly drying signal of around 25-50\% across the North Island, as well as across large regions of the South Island in CanESM5, ACCESS-ESM1-5 and the dynamically downscaled ensemble. Changes in the duration of the median longest drought differ far more, and are analogous to the changes seen for all droughts shown in Figure \ref{cc-signals}, with droughts in the dynamically downscaled ensemble becoming 3-5 months longer across the North Island, and north east of the South Island, and 1-2 months longer over the same area in ACCESS-ESM1-5. The median longest drought decrease in duration across the majority of the country by 2-3 months in CanESM5, with more muted changes across the other CMIP6 GCMs.  \\

The worst case scenario for the future changes in the median longest drought among members, is that occurring in the dynamically downscaled ensemble. Across some regions of the the North Island if droughts were to get up to 50\% more intense, and last 5 month longer it is possible that a median drought within a 50 year period in the future could have 50\% less rain than average over a 12 month period. This precipitation deficit over such a long duration is particularly significant due to the low precipitation variability typical of New Zealand's maritime climate. Such a drought would have extreme and widespread impacts across all primary sectors, worse than the effects of recent events in 1997/98, 2007/08 or 2012/13, each of which resulted in billions of dollars of economic losses \citep{kamber2013drying,ford2015nino}.\\ 

\begin{figure}
    \centering
    \makebox[\textwidth][c]{\includegraphics[width=1.2\textwidth]{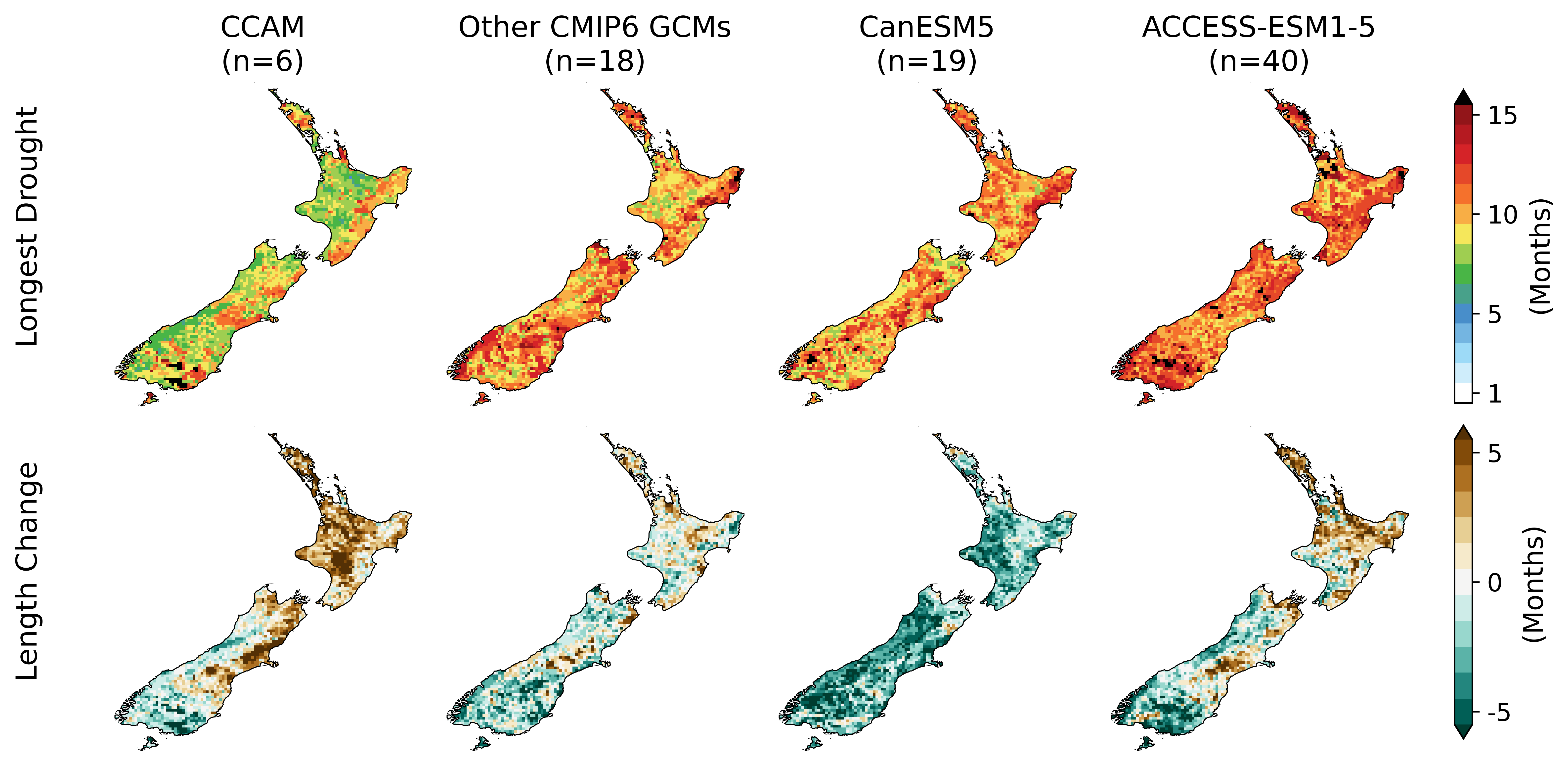}} 
    \caption{First Row: Longest drought across all members of each ensemble at each grid-point within an extended historical period (1965-2014). Second Row: The change in longest drought across all members in an extended future period (2050-2099) under an SSP3-7.0 scenario. Here, ensembles are arranged from smallest to largest to reflect the increased likelihood of observing more extreme events in a larger ensemble. }
    \label{longest_drought}
\end{figure}

With the effects of extreme droughts in mind, it is useful to consider the longest possible drought which could conceivably occur in today's climate, and how such droughts will change in the future under a high emission SSP3-7.0 scenario. Figure \ref{longest_drought} shows the longest drought at each grid-point across all members within an extended historical period (1965-2014), as well as the relative length of the longest drought in an extended future period (2050-2099). Here, the length of the longest drought observed in each ensemble is highly dependent on the size of the ensemble itself, which is consistent with the increase in sample size that large ensembles provide \citep{suarez2018internal,fischer2013robust,haugen2018estimating}. We show this is true in the case of this dataset using subsamples of the ACCESS-ESM1-5 ensemble which we show in the supplementary material. The longest droughts seen in the largest ensemble ACCESS-ESM1-5 (n=40), are on the order of 11-12 months long across the majority of the country, with some grid-cells having droughts 20 months in length. The maximum length of droughts seen in recent observational record (VCSN, 1972-2020) are only on the order of 5-7 months across most of the country (see Figure S28 of the supplementary material). A drought 20 months in long in the current climate would have catastrophic consequences for the primary sector as well as energy and water security nationwide \citep{hendy2018drought}. The change in length of the longest drought in the future period is again unique to each ensemble, with all ensembles projecting decreases in the south west of the country, and the dynamically downscaled ensemble, as well as the ACCESS-ESM1-5 ensemble showing significant increases in duration over the north of the country. \\ 

\begin{figure}
    \centering
    \makebox[\textwidth][c]{\includegraphics[width=1\textwidth]{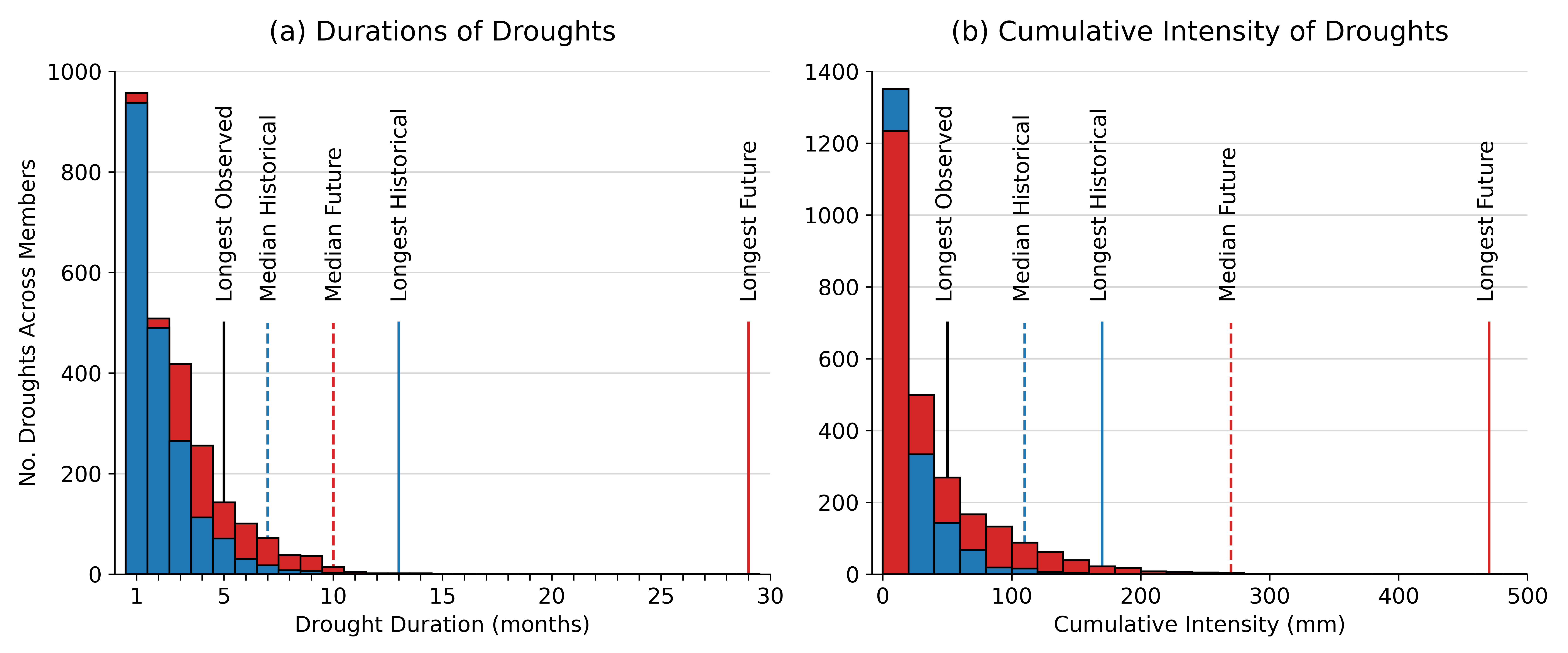}} 
    \caption{Histograms of drought duration (a), and cumulative drought intensity (b), at a location in the far north of the North Island (34.68°S,173.12°E), across all members of ACCESS-ESM1-5. Blue bars denote data from the extended historical period (1965-2014), red bars denote data from the extended future period 2050-2099. This location contains the longest drought within the entire AI-downscaled dataset, 29 months, within one members future period. Vertical lines show the duration and intensity of the median longest drought among ACCESS-ESM1-5 members, and the longest drough among ACCESS-ESM1-5 members at this location, in both extended historical and future periods. Black lines denote the duration and intensity of the longest drought seen in the observed VCSN dataset across 1972-2020.}
    \label{hist}
\end{figure}

Within the extended future period of one member of ACCESS-ESM1-5 (r18i1p1f1), at one particular grid-point (the far north of the North Island (34.68°S,173.12°E)), an event with a duration of 29 months is present. This is the longest drought we see anywhere across all ensembles, with a duration that is completely unprecedented in both current and future climates. We contextualize the relative rarity of this unprecedented drought against all droughts present in the ACCESS-ESM1-5 ensemble and VCSN observations at this particular location in Figure \ref{hist}. In panel (a) we see that this particular event is over twice as long as the longest event present in the historical period, over four times as long as the median longest historical drought across members, and nearly six times as long as any drought seen in the recent observational record. This drought is not only long, but incredibly intense, with a cumulative intensity of 478 mm below the 15th percentile of precipitation over this 29 month period, which is approximately 48\% less precipitation than would occur over this span in the historical period.  \\ 

It is important to keep in mind that the extreme droughts presented in Figures \ref{longest_drought} and \ref{hist} represent only the most severe droughts which arise over thousands of years of simulations (2000 in both historical and future periods in the case of the ACCESS-ESM1-5 ensemble). These events serve as an upper bound of drought severity which could occur in both the current and a future climate under an SSP3-7.0 scenario, and not one that would be typical to see under any circumstances. A more reasonable estimate of a severe drought likely to occur in a 50 year stretch in both current and future climates would be the median longest events among members depicted in Figure \ref{median_drought}.

\section{Discussion}\label{sec12}

One of the frequently raised issues regarding the use of AI-downscaling algorithms to produce climate projections is their ability to generalize beyond their training dataset, particularly for extreme events \citep{doury2024suitability,addison2025machinelearningemulationprecipitation,PotentialforMachineLearningEmulatorstoAugmentRegionalClimateSimulationsinProvisionofLocalClimateChangeInformation}. However, the AI-RCME used here has been previously demonstrated the ability to capture climate change signals for both climatic means, and extremes (RX1day, RX10yea), in GCM data not seen during training \citep{rampal2025b}. Here, we demonstrate this to be true also, at the other end of the precipitation spectrum. This ability to correctly generalize meteorological droughts is in theory an innate strength of any AI-RCME that is able to accurately capture mean precipitation statistics of other GCMs, as well as precipitation extremes, as these could constitute drought breaking events based on our definitions.  \\

    Something to consider about the use of AI-downscaling to investigate drought, is the downscaling procedure itself is critically important to correctly produce certain metrics, for example soil moisture. In the dataset presented in \cite{rampal2025b}, and used here precipitation and temperature are downscaled independently of each other, with each day being downscaled independently of any other timestep. This means that the land surface has no persistent memory of the soil moisture state, and thus the impacts of soil moisture on near surface temperatures or precipitation generation is not directly captured. This would imply key processes such as drought/temperature intensification through land-atmosphere feedbacks would not be present either, reducing the overall realism of soil-moisture droughts simulated with such a model. This is less important for meteorological drought as examined here - but would be key for representing other types of drought (e.g. agricultural, hydrological).  Future AI-downscaling approaches would need to implement a form of interactive land-surface, with a persistent memory of the soil moisture state to correctly simulate these effects. Moreover, the independent simulation of each day has implications for the simulation of persistent synoptic systems. Architectures with a ``memory'' of previous states (e.g., autoregressive approaches), could possibly better represent these systems, and represent an important gap in the literature. Furthermore, any future improvements in RCM emulation, not limited to incorporating land–atmosphere coupling should be benchmarked against our results, which we have shown to be robust against physics-based dynamical downscaling. Interpreting these advances may help further reduce some aspects of AI-model-uncertainty. \\ 

The selection of any drought metric must always come with an awareness of how meaningfully that metric represents drought related impacts, and how confident we are in the modeling tools available to simulate droughts for the right reasons. This includes being able to identify the physical processes which govern future changes in droughts in a warming world \citep{erian2021gar}. Here, we are constrained by the number of available outputs of the AI-RCME, and the configuration of the AI itself as discussed in the previous paragraph. Thus we are only able to investigate meteorological droughts, rather than other possibly more impactful metrics which would require a larger number of downscaled variables to calculate, and a different downscaling architecture. This being said, the meteorological drought metrics we do choose to use have been shown to have highly robust agreement across CMIP5 and CMIP6 GCMs \citep{ukkola2020}, more so than changes in mean precipitation \citep{collins2013long}, a key point considering the large number of different GCMs used in this study.  \\

During our analysis we compared results derived from CMIP6 GCM downscaled using an AI-RCME to the ensemble of RCMs dynamically downscaled with CCAM. The GCMs downscaled with the AI-RCME (n=20) were chosen on the basis of data availability, while the GCMs chosen for dynamical downscaling (n=6) were chosen based on their ability to simulate New Zealand historical climate (see section \ref{RCM_data}). Some of the GCMs downscaled using the AI-RCME do quite a poor job at simulating New Zealand's historical climate across both large-scale and small-scale features, an evaluation of this is shown in Figures S1 and S2 within the supplementary material of \cite{gibson2024dynamical}, and the supplementary material of \cite{rampal2025b}. It is important to address that while some GCMs may better represent New Zealand's historical climate than others, that their performance in a future period may not necessarily be physically accurate or more realistic than other models. This being the premise of model uncertainty \citep{hawkins2009potential}. Methods such as emergent constraints can also be used to constrain future projections using statistical relationships between aspects of current climate \citep{hall2019progressing,sherwood2020assessment}, however this goes well beyond the scope of our study. \\

\section{Conclusion}\label{sec13}

In this paper, we have used the outputs of an AI-RCME to produce future projections of meteorological drought across New Zealand under a high emission SSP3-7.0 scenario. The AI-downscaled dataset is far larger than could be otherwise achieved using conventional dynamical downscaling, consisting of 20 GCMs, including one 40 member initial condition ensemble, and one 19 member initial condition ensemble. The unprecedented size of this over 15000 year dataset allows additional insights into changes of future droughts not possible with smaller dynamically downscaled ensembles.    \\

We first demonstrate that the AI-RCME correctly emulates the downscaling ability of the CCAM RCM, across annual and seasonal precipitation, as well as the drought metrics introduced in Section \ref{Drought_Metrics}. The AI-RCME adds a similar amount of value to the CCAM RCM for annual and seasonal precipitation, as well as drought intensity. Comparisons of drought durations and frequencies to observations are obscured by natural variability. The AI-RCME successfully reproduces the climate change signal for precipitation as shown by \cite{rampal2025b} (Figures S20 and S21). This precipitation climate change signal closely resembling that of the dynamically downscaled model with some characteristics of the host GCM. We find similar behavior for the drought metrics we consider, with their climate change signals also resembling a midpoint between the host GCM and dynamically downscaled RCM. \\

We then used this dataset to address an open question on the sign changes of DJF precipitation across New Zealand. We find that existing dynamically downscaled projections may be too confident in projecting this sign change, as there are large deviations in future precipitation changes driven by internal variability and model uncertainty, which the dynamically downscaled ensemble does not directly account for. Assessments of scenario uncertainty were not performed in this study, as only an SSP3-7.0 future scenario was produced for all the models assessed here. This will be a focus of future works as these downscaled simulations become available. \\

Even in the presence of internal variability, projections of of drought intensity are robust across all dynamically downscaled and AI-RCME projections with future droughts to become approximately 25-50\% more intense across the majority of the country. These increases are robust even in the presence of increases in mean precipitation, with increases in drought intensity being exclusively driven by increases in precipitation variability across many regions across the ensembles. Changes in drought durations generally correspond to changes in mean precipitation which differ significantly between ensembles. Regions with weaker increases in mean precipitation often had highly uncertain future changes in drought duration and frequency.  \\ 

With the increased sample size provided by the large downscaled ensembles we were able to gain insights around the longest droughts possible in the current climate, as well as under a high emission SSP3-7.0 scenario. The
median longest among members seen in the current climate is approximately 6-7 months long across the entirety of the country with these drought resulting in 40\% less precipitation over this period. Under a high emission SSP3-7.0 scenario this median longest drought could approach 12 months long, with 50\% less rain falling over this period. We estimate that the longest possible drought in the current climate is approximately 12 months long for the majority of the country. In the future this could increase by over 5 months in some regions. We found a particularly interesting 29 month long drought in the far north of the North Island in the future, with only 48\% of the average precipitation falling over this period. \\ 

Here, we have demonstrated the value of AI-RCME to enhance our knowledge of regional climatic changes with a particular focus on meteorological drought. We envision the framework used here and in \cite{rampal2025b}: leveraging regional climate insights from data produced using AI-RCME can be implemented by other countries with limited downscaling resources to produce more robust estimates of changes at regional scales, and to explore and quantify different sources of uncertainty. The use of large-ensembles produced using this framework can also aid in adaptation and mitigation efforts as they allow for risk-based assessments \citep{sutton2019climate}, which can be highly useful when projections are uncertain \citep{sherwood2024uncertain,shepherd2018storylines}.  \\

\backmatter

\bmhead{Supplementary information}

Please see the supplementary material for additional analysis. 

\bmhead{Acknowledgements}

HL and LH acknowledge funding from the Royal Society of New Zealand via the Marsden Fund (Grant ID: MFP-UOW2307) and New Zealand’s Ministry for Business, Innovation \& Employment (MBIE) via their Endeavour Fund Smart Ideas Fund (Grant ID: UOWX2302). NR acknowledges the support
of the Australian Research Council Centre of Excellence for Climate Extremes (CLEX; CE170100023). NR and PG acknowledge funding from the New Zealand MBIE Endeavour Smart Ideas Fund (C01X2202). CH acknowledges funding from the Australian Research Council Centre of Excellence for the Weather of the 21st Century (CE230100012). AU acknowledges the funding of the Australian Research Council Discovery Early Career Researcher Award
Project (DE200100086). NM acknowledges the funding of the Australian Research Council Discovery Early Career Researcher Award
Project (DE230100315). \\

HL acknowledges Andrew King who provided helpful discussions on research direction.  

\bmhead{Data Availability}

The CCAM RCM climate projection ensemble used in this work was produced by NIWA \citep{Gibson2023,gibson2024dynamical}. This CCAM RCM downscaled data is freely available at https://climatedata.environment.govt.nz/. \\

The AI-RCME code and supporting sample datasets are available here: https://zenodo.org/records/13755688. \\ 

\noindent GCM data downscaled in this analysis is available from ESGF: https://esgf.github.io/nodes.html

\bmhead{Code Availability}

The code for training the RCM emulator is available at: https://github.com/nram812/On-the-Extrapolation-of-Generative-Adversarial-Networks-for-downscaling-precipitation-extremes. \\

The AI-RCME code and supporting sample datasets are available here: https://zenodo.org/records/13755688. \\ 

\bmhead{Competing Interests}

The authors declare no competing interests.

\bmhead{Author Contributions}

HL: conceptualization, methodology, analysis, visualization, writing/editing. NR: produced AI dataset, conceptualization, methodology, writing/editing. PG, LH, CH, AU, NM: conceptualization, methodology, writing/editing.








\bibliography{sn-bibliography.bib}

\begin{thebibliography}{68}
\providecommand{\natexlab}[1]{#1}
\providecommand{\url}[1]{{#1}}
\providecommand{\urlprefix}{URL }
\providecommand{\doi}[1]{\url{https://doi.org/#1}}
\providecommand{\eprint}[2][]{\url{#2}}
 \bibcommenthead

\bibitem[{Aalbers et~al.(2018)Aalbers, Lenderink, van Meijgaard, and van~den Hurk}]{aalbers2018local}
Aalbers EE, Lenderink G, van Meijgaard E, et~al (2018) Local-scale changes in mean and heavy precipitation in western europe, climate change or internal variability? Climate Dynamics 50:4745--4766

\bibitem[{Addison et~al.(2025)Addison, Kendon, Ravuri, Aitchison, and Watson}]{addison2025machinelearningemulationprecipitation}
Addison H, Kendon E, Ravuri S, et~al (2025) Machine learning emulation of precipitation from km-scale regional climate simulations using a diffusion model. \urlprefix\url{https://arxiv.org/abs/2407.14158}, {\href{https://arxiv.org/abs/2407.14158}{{arXiv:2407.14158}}}

\bibitem[{Bailie et~al.(2024)Bailie, Koh, Rampal, and Gibson}]{bailie2024quantile}
Bailie T, Koh YS, Rampal N, et~al (2024) Quantile-regression-ensemble: A deep learning algorithm for downscaling extreme precipitation. In: Proceedings of the aaai conference on artificial intelligence, pp 21914--21922

\bibitem[{Ba{\~n}o-Medina et~al.(2024)Ba{\~n}o-Medina, Iturbide, Fern{\'a}ndez, and Guti{\'e}rrez}]{bano2024transferability}
Ba{\~n}o-Medina J, Iturbide M, Fern{\'a}ndez J, et~al (2024) Transferability and explainability of deep learning emulators for regional climate model projections: Perspectives for future applications. Artificial Intelligence for the Earth Systems 3(4):e230099

\bibitem[{Bart{\'o}k et~al.(2017)Bart{\'o}k, Wild, Folini, L{\"u}thi, Kotlarski, Sch{\"a}r, Vautard, Jerez, and Imecs}]{bartok2017projected}
Bart{\'o}k B, Wild M, Folini D, et~al (2017) Projected changes in surface solar radiation in cmip5 global climate models and in euro-cordex regional climate models for europe. Climate dynamics 49(7):2665--2683

\bibitem[{Bengtsson and Hodges(2019)}]{bengtsson2019can}
Bengtsson L, Hodges KI (2019) Can an ensemble climate simulation be used to separate climate change signals from internal unforced variability? Climate Dynamics 52(5):3553--3573

\bibitem[{Bo{\'e} et~al.(2023)Bo{\'e}, Mass, and Deman}]{boe2023simple}
Bo{\'e} J, Mass A, Deman J (2023) A simple hybrid statistical--dynamical downscaling method for emulating regional climate models over western europe. evaluation, application, and role of added value? Climate Dynamics 61(1):271--294

\bibitem[{Chadwick et~al.(2011)Chadwick, Coppola, and Giorgi}]{chadwick2011artificial}
Chadwick R, Coppola E, Giorgi F (2011) An artificial neural network technique for downscaling gcm outputs to rcm spatial scale. Nonlinear Processes in Geophysics 18(6):1013--1028

\bibitem[{Collins et~al.(2018)Collins, Montgomery, and Zammit}]{collins2018hydrological}
Collins D, Montgomery K, Zammit C (2018) Hydrological projections for new zealand rivers under climate change. Ministry for the Environment, NZ

\bibitem[{Collins et~al.(2013)Collins, Knutti, Arblaster, Dufresne, Fichefet, Friedlingstein, Gao, Gutowski, Johns, Krinner et~al.}]{collins2013long}
Collins M, Knutti R, Arblaster J, et~al (2013) Long-term climate change: projections, commitments and irreversibility. In: Climate change 2013-The physical science basis: Contribution of working group I to the fifth assessment report of the intergovernmental panel on climate change. Cambridge University Press, p 1029--1136

\bibitem[{Deser et~al.(2012)Deser, Phillips, Bourdette, and Teng}]{deser2012uncertainty}
Deser C, Phillips A, Bourdette V, et~al (2012) Uncertainty in climate change projections: the role of internal variability. Climate dynamics 38:527--546

\bibitem[{Deser et~al.(2014)Deser, Phillips, Alexander, and Smoliak}]{deser2014}
Deser C, Phillips AS, Alexander MA, et~al (2014) Projecting north american climate over the next 50 years: Uncertainty due to internal variability. Journal of Climate 27(6):2271 -- 2296. \doi{10.1175/JCLI-D-13-00451.1}, \urlprefix\url{https://journals.ametsoc.org/view/journals/clim/27/6/jcli-d-13-00451.1.xml}

\bibitem[{Di~Virgilio et~al.(2020)Di~Virgilio, Evans, Di~Luca, Grose, Round, and Thatcher}]{di2020realised}
Di~Virgilio G, Evans JP, Di~Luca A, et~al (2020) Realised added value in dynamical downscaling of australian climate change. Climate Dynamics 54:4675--4692

\bibitem[{Douris and Kim(2021)}]{douris2021atlas}
Douris J, Kim G (2021) The atlas of mortality and economic losses from weather, climate and water extremes (1970-2019). N/A

\bibitem[{Doury et~al.(2023)Doury, Somot, Gadat, Ribes, and Corre}]{doury2023regional}
Doury A, Somot S, Gadat S, et~al (2023) Regional climate model emulator based on deep learning: Concept and first evaluation of a novel hybrid downscaling approach. Climate Dynamics 60(5):1751--1779

\bibitem[{Doury et~al.(2024)Doury, Somot, and Gadat}]{doury2024suitability}
Doury A, Somot S, Gadat S (2024) On the suitability of a convolutional neural network based rcm-emulator for fine spatio-temporal precipitation. Climate Dynamics 62(9):8587--8613

\bibitem[{Erian et~al.(2021)Erian, Pulwarty, Vogt, AbuZeid, Bert, Bruntrup, El-Askary, de~Estrada, Gaupp, Grundy et~al.}]{erian2021gar}
Erian W, Pulwarty R, Vogt J, et~al (2021) GAR special report on drought 2021. United Nations Office for Disaster Risk Reduction (UNDRR)

\bibitem[{Eyring et~al.(2016)Eyring, Bony, Meehl, Senior, Stevens, Stouffer, and Taylor}]{Eyring2016}
Eyring V, Bony S, Meehl GA, et~al (2016) Overview of the coupled model intercomparison project phase 6 (cmip6) experimental design and organization. Geoscientific Model Development 9(5):1937--1958. \doi{10.5194/gmd-9-1937-2016}, \urlprefix\url{https://gmd.copernicus.org/articles/9/1937/2016/}

\bibitem[{Falster et~al.(2024)Falster, Wright, Abram, Ukkola, and Henley}]{Falster2024}
Falster GM, Wright NM, Abram NJ, et~al (2024) Potential for historically unprecedented australian droughts from natural variability and climate change. Hydrology and Earth System Sciences 28(6):1383--1401. \doi{10.5194/hess-28-1383-2024}, \urlprefix\url{https://hess.copernicus.org/articles/28/1383/2024/}

\bibitem[{Fischer et~al.(2013)Fischer, Beyerle, and Knutti}]{fischer2013robust}
Fischer EM, Beyerle U, Knutti R (2013) Robust spatially aggregated projections of climate extremes. Nature Climate Change 3(12):1033--1038

\bibitem[{Ford and Wood(2015)}]{ford2015nino}
Ford D, Wood A (2015) El ni{\~n}o and its impact on the new zealand economy. Tech. rep., Reserve Bank of New Zealand

\bibitem[{Gibson et~al.(2016)Gibson, Perkins-Kirkpatrick, and Renwick}]{gibson2016}
Gibson PB, Perkins-Kirkpatrick SE, Renwick JA (2016) Projected changes in synoptic weather patterns over new zealand examined through self-organizing maps. International Journal of Climatology 36(12):3934--3948. \doi{https://doi.org/10.1002/joc.4604}, \urlprefix\url{https://rmets.onlinelibrary.wiley.com/doi/abs/10.1002/joc.4604}, {\href{https://arxiv.org/abs/https://rmets.onlinelibrary.wiley.com/doi/pdf/10.1002/joc.4604}{{https://rmets.onlinelibrary.wiley.com/doi/pdf/10.1002/joc.4604}}}

\bibitem[{Gibson et~al.(2023)Gibson, Stone, Thatcher, Broadbent, Dean, Rosier, Stuart, and Sood}]{Gibson2023}
Gibson PB, Stone D, Thatcher M, et~al (2023) High-resolution ccam simulations over new zealand and the south pacific for the detection and attribution of weather extremes. Journal of Geophysical Research: Atmospheres 128(14):e2023JD038530. \doi{https://doi.org/10.1029/2023JD038530}, \urlprefix\url{https://agupubs.onlinelibrary.wiley.com/doi/abs/10.1029/2023JD038530}, e2023JD038530 2023JD038530, {\href{https://arxiv.org/abs/https://agupubs.onlinelibrary.wiley.com/doi/pdf/10.1029/2023JD038530}{{https://agupubs.onlinelibrary.wiley.com/doi/pdf/10.1029/2023JD038530}}}

\bibitem[{Gibson et~al.(2024{\natexlab{a}})Gibson, Rampal, Dean, and Morgenstern}]{gibson2024storylines}
Gibson PB, Rampal N, Dean SM, et~al (2024{\natexlab{a}}) Storylines for future projections of precipitation over new zealand in cmip6 models. Journal of Geophysical Research: Atmospheres 129(5):e2023JD039664

\bibitem[{Gibson et~al.(2024{\natexlab{b}})Gibson, Stuart, Sood, Stone, Rampal, Lewis, Broadbent, Thatcher, and Morgenstern}]{gibson2024dynamical}
Gibson PB, Stuart S, Sood A, et~al (2024{\natexlab{b}}) Dynamical downscaling cmip6 models over new zealand: added value of climatology and extremes. Climate Dynamics pp 1--27

\bibitem[{Gibson et~al.(2025)Gibson, Broadbent, Stuart, Lewis, Campbell, Rampal, Harrington, and Williams}]{gibson2025downscaled}
Gibson PB, Broadbent AM, Stuart SJ, et~al (2025) Downscaled cmip6 future climate projections for new zealand: climatology and extremes. Weather and Climate Extremes p 100784

\bibitem[{Hall et~al.(2019)Hall, Cox, Huntingford, and Klein}]{hall2019progressing}
Hall A, Cox P, Huntingford C, et~al (2019) Progressing emergent constraints on future climate change. Nature Climate Change 9(4):269--278

\bibitem[{Harrington et~al.(2016)Harrington, Gibson, Dean, Mitchell, Rosier, and Frame}]{harrington2016}
Harrington LJ, Gibson PB, Dean SM, et~al (2016) Investigating event-specific drought attribution using self-organizing maps. Journal of Geophysical Research: Atmospheres 121(21):12,766--12,780. \doi{https://doi.org/10.1002/2016JD025602}, \urlprefix\url{https://agupubs.onlinelibrary.wiley.com/doi/abs/10.1002/2016JD025602}, {\href{https://arxiv.org/abs/https://agupubs.onlinelibrary.wiley.com/doi/pdf/10.1002/2016JD025602}{{https://agupubs.onlinelibrary.wiley.com/doi/pdf/10.1002/2016JD025602}}}

\bibitem[{Haugen et~al.(2018)Haugen, Stein, Moyer, and Sriver}]{haugen2018estimating}
Haugen MA, Stein ML, Moyer EJ, et~al (2018) Estimating changes in temperature distributions in a large ensemble of climate simulations using quantile regression. Journal of CLIMATE 31(20):8573--8588

\bibitem[{Hawkins and Sutton(2009)}]{hawkins2009potential}
Hawkins E, Sutton R (2009) The potential to narrow uncertainty in regional climate predictions. Bulletin of the American Meteorological Society 90(8):1095--1108

\bibitem[{Hendy et~al.(2018)Hendy, Kerr, Halliday, Owen, Ausseil, Bell, Deans, Dickie, Hale, Hale et~al.}]{hendy2018drought}
Hendy J, Kerr S, Halliday A, et~al (2018) Drought and climate change adaptation: impacts and projections. Motu Economic and Public Policy Research Wellington, New Zealand

\bibitem[{Ji et~al.(2024)Ji, Fu, Liu, Huang, and Tan}]{ji2024uncertainty}
Ji Y, Fu J, Liu B, et~al (2024) Uncertainty separation of drought projection in the 21st century using smiles and cmip6. Journal of Hydrology 628:130497

\bibitem[{Kamber et~al.(2013)Kamber, McDonald, Price et~al.}]{kamber2013drying}
Kamber G, McDonald C, Price G, et~al (2013) Drying out: Investigating the economic effects of drought in new zealand. Tech. rep., Reserve Bank of New Zealand Wellington

\bibitem[{Kendon et~al.(2025)Kendon, Addison, Doury, Somot, Watson, Booth, Coppola, Gutiérrez, Murphy, and Scullion}]{PotentialforMachineLearningEmulatorstoAugmentRegionalClimateSimulationsinProvisionofLocalClimateChangeInformation}
Kendon EJ, Addison H, Doury A, et~al (2025) Potential for machine learning emulators to augment regional climate simulations in provision of local climate change information. Bulletin of the American Meteorological Society 106(6):E1175 -- E1203. \doi{10.1175/BAMS-D-24-0114.1}, \urlprefix\url{https://journals.ametsoc.org/view/journals/bams/106/6/BAMS-D-24-0114.1.xml}

\bibitem[{Leduc et~al.(2019)Leduc, Mailhot, Frigon, Martel, Ludwig, Brietzke, Gigu{\`e}re, Brissette, Turcotte, Braun et~al.}]{leduc2019climex}
Leduc M, Mailhot A, Frigon A, et~al (2019) The climex project: A 50-member ensemble of climate change projections at 12-km resolution over europe and northeastern north america with the canadian regional climate model (crcm5). Journal of Applied Meteorology and Climatology 58(4):663--693

\bibitem[{Lesk et~al.(2016)Lesk, Rowhani, and Ramankutty}]{lesk2016influence}
Lesk C, Rowhani P, Ramankutty N (2016) Influence of extreme weather disasters on global crop production. Nature 529(7584):84--87

\bibitem[{Lewis et~al.(2025)Lewis, Harrington, Gibson, and Rampal}]{Lewis2025}
Lewis H, Harrington LJ, Gibson PB, et~al (2025) Storylines of future drought in the face of uncertain rainfall projections: a new zealand case study. EGUsphere 2025:1--16. \doi{10.5194/egusphere-2025-1247}, \urlprefix\url{https://egusphere.copernicus.org/preprints/2025/egusphere-2025-1247/}

\bibitem[{Maher et~al.(2021)Maher, Milinski, and Ludwig}]{maher2021large}
Maher N, Milinski S, Ludwig R (2021) Large ensemble climate model simulations: introduction, overview, and future prospects for utilising multiple types of large ensemble. Earth System Dynamics 12(2):401--418

\bibitem[{McGregor and Dix(2008)}]{mcgregor2008updated}
McGregor JL, Dix MR (2008) An updated description of the conformal-cubic atmospheric model. In: High resolution numerical modelling of the atmosphere and ocean. Springer, p 51--75

\bibitem[{Mullan et~al.(2018)Mullan, Abha, and Stuart}]{mullen2018}
Mullan B, Abha S, Stuart S (2018) Climate change projections for new zealand: Atmospheric projections based on simulations undertaken for the ipcc 5th assessment. National Institute of Water \& Atmospheric Research

\bibitem[{Naumann et~al.(2021)Naumann, Cammalleri, Mentaschi, and Feyen}]{naumann2021increased}
Naumann G, Cammalleri C, Mentaschi L, et~al (2021) Increased economic drought impacts in europe with anthropogenic warming. Nature Climate Change 11(6):485--491

\bibitem[{Rampal et~al.(2024{\natexlab{a}})Rampal, Hobeichi, Gibson, Ba{\~n}o-Medina, Abramowitz, Beucler, Gonz{\'a}lez-Abad, Chapman, Harder, and Guti{\'e}rrez}]{rampal2024enhancing}
Rampal N, Hobeichi S, Gibson PB, et~al (2024{\natexlab{a}}) Enhancing regional climate downscaling through advances in machine learning. Artificial Intelligence for the Earth Systems 3(2):230066

\bibitem[{Rampal et~al.(2024{\natexlab{b}})Rampal, Hobeichi, Gibson, Baño-Medina, Abramowitz, Beucler, González-Abad, Chapman, Harder, and Gutiérrez}]{rampal2024}
Rampal N, Hobeichi S, Gibson PB, et~al (2024{\natexlab{b}}) Enhancing regional climate downscaling through advances in machine learning. Artificial Intelligence for the Earth Systems 3(2):230066. \doi{10.1175/AIES-D-23-0066.1}, \urlprefix\url{https://journals.ametsoc.org/view/journals/aies/3/2/AIES-D-23-0066.1.xml}

\bibitem[{Rampal et~al.(2025{\natexlab{a}})Rampal, Gibson, Sherwood, Abramowitz, and Hobeichi}]{rampal2025reliable}
Rampal N, Gibson PB, Sherwood S, et~al (2025{\natexlab{a}}) A reliable generative adversarial network approach for climate downscaling and weather generation. Journal of Advances in Modeling Earth Systems 17(1):e2024MS004668

\bibitem[{Rampal et~al.(2025{\natexlab{b}})Rampal, Gibson, Sherwood, Queen, Lewis, and Abramowitz}]{rampal2025b}
Rampal N, Gibson PB, Sherwood SC, et~al (2025{\natexlab{b}}) Downscaling with ai reveals the large role of internal variability in fine-scale projections of climate extremes. N/A \urlprefix\url{https://arxiv.org/abs/2507.06527}, {\href{https://arxiv.org/abs/2507.06527}{{arXiv:2507.06527}}} {[physics.ao-ph]}

\bibitem[{Renwick et~al.(1998)Renwick, Katzfey, Nguyen, and McGregor}]{Renwick1998}
Renwick JA, Katzfey JJ, Nguyen KC, et~al (1998) Regional model simulations of new zealand climate. Journal of Geophysical Research: Atmospheres 103(D6):5973--5982. \doi{https://doi.org/10.1029/97JD02939}, \urlprefix\url{https://agupubs.onlinelibrary.wiley.com/doi/abs/10.1029/97JD02939}, {\href{https://arxiv.org/abs/https://agupubs.onlinelibrary.wiley.com/doi/pdf/10.1029/97JD02939}{{https://agupubs.onlinelibrary.wiley.com/doi/pdf/10.1029/97JD02939}}}

\bibitem[{Ronneberger et~al.(2015)Ronneberger, Fischer, and Brox}]{ronneberger2015u}
Ronneberger O, Fischer P, Brox T (2015) U-net: Convolutional networks for biomedical image segmentation. In: International Conference on Medical image computing and computer-assisted intervention, Springer, pp 234--241

\bibitem[{Rummukainen(2016)}]{rummukainen2016added}
Rummukainen M (2016) Added value in regional climate modeling. Wiley Interdisciplinary Reviews: Climate Change 7(1):145--159

\bibitem[{Samaniego et~al.(2018)Samaniego, Thober, Kumar, Wanders, Rakovec, Pan, Zink, Sheffield, Wood, and Marx}]{samaniego2018anthropogenic}
Samaniego L, Thober S, Kumar R, et~al (2018) Anthropogenic warming exacerbates european soil moisture droughts. Nature Climate Change 8(5):421--426

\bibitem[{Shepherd et~al.(2018)Shepherd, Boyd, Calel, Chapman, Dessai, Dima-West, Fowler, James, Maraun, Martius et~al.}]{shepherd2018storylines}
Shepherd TG, Boyd E, Calel RA, et~al (2018) Storylines: an alternative approach to representing uncertainty in physical aspects of climate change. Climatic change 151:555--571

\bibitem[{Sherwood et~al.(2024)Sherwood, Hegerl, Braconnot, Friedlingstein, Goelzer, Harris, Holland, Kim, Mitchell, Naish et~al.}]{sherwood2024uncertain}
Sherwood S, Hegerl G, Braconnot P, et~al (2024) Uncertain pathways to a future safe climate. Earth's Future 12(6):e2023EF004297

\bibitem[{Sherwood et~al.(2020)Sherwood, Webb, Annan, Armour, Forster, Hargreaves, Hegerl, Klein, Marvel, Rohling et~al.}]{sherwood2020assessment}
Sherwood SC, Webb MJ, Annan JD, et~al (2020) An assessment of earth's climate sensitivity using multiple lines of evidence. Reviews of Geophysics 58(4):e2019RG000678

\bibitem[{Singh et~al.(2022)Singh, Ashfaq, Skinner, Anderson, Mishra, and Singh}]{singh2022enhanced}
Singh J, Ashfaq M, Skinner CB, et~al (2022) Enhanced risk of concurrent regional droughts with increased enso variability and warming. Nature Climate Change 12(2):163--170

\bibitem[{Sousa et~al.(2018)Sousa, Blamey, Reason, Ramos, and Trigo}]{sousa2018day}
Sousa PM, Blamey RC, Reason CJ, et~al (2018) The ‘day zero’cape town drought and the poleward migration of moisture corridors. Environmental Research Letters 13(12):124025

\bibitem[{Stevenson et~al.(2022)Stevenson, Coats, Touma, Cole, Lehner, Fasullo, and Otto-Bliesner}]{stevenson2022twenty}
Stevenson S, Coats S, Touma D, et~al (2022) Twenty-first century hydroclimate: A continually changing baseline, with more frequent extremes. Proceedings of the National Academy of Sciences 119(12):e2108124119

\bibitem[{Suarez-Gutierrez et~al.(2018)Suarez-Gutierrez, Li, M{\"u}ller, and Marotzke}]{suarez2018internal}
Suarez-Gutierrez L, Li C, M{\"u}ller WA, et~al (2018) Internal variability in european summer temperatures at 1.5 c and 2 c of global warming. Environmental Research Letters 13(6):064026

\bibitem[{Sutton(2019)}]{sutton2019climate}
Sutton RT (2019) Climate science needs to take risk assessment much more seriously. Bulletin of the American Meteorological Society 100(9):1637--1642

\bibitem[{Tait et~al.(2006)Tait, Henderson, Turner, and Zheng}]{tait2006}
Tait A, Henderson R, Turner R, et~al (2006) Thin plate smoothing spline interpolation of daily rainfall for new zealand using a climatological rainfall surface. International Journal of Climatology 26(14):2097--2115. \doi{https://doi.org/10.1002/joc.1350}, \urlprefix\url{https://rmets.onlinelibrary.wiley.com/doi/abs/10.1002/joc.1350}, {\href{https://arxiv.org/abs/https://rmets.onlinelibrary.wiley.com/doi/pdf/10.1002/joc.1350}{{https://rmets.onlinelibrary.wiley.com/doi/pdf/10.1002/joc.1350}}}

\bibitem[{Tait et~al.(2012)Tait, Sturman, and Clark}]{tait2012}
Tait A, Sturman J, Clark M (2012) An assessment of the accuracy of interpolated daily rainfall for new zealand. Journal of Hydrology (New Zealand) 51(1):25--44. \urlprefix\url{http://www.jstor.org/stable/43944886}

\bibitem[{Trenberth(1976)}]{trenberth1976fluctuations}
Trenberth KE (1976) Fluctuations and trends in indices of the southern hemispheric circulation. Quarterly Journal of the Royal Meteorological Society 102(431):65--75

\bibitem[{Trenberth et~al.(2014)Trenberth, Dai, Van Der~Schrier, Jones, Barichivich, Briffa, and Sheffield}]{trenberth2014global}
Trenberth KE, Dai A, Van Der~Schrier G, et~al (2014) Global warming and changes in drought. Nature Climate Change 4(1):17--22

\bibitem[{von Trentini et~al.(2019)von Trentini, Leduc, and Ludwig}]{von2019assessing}
von Trentini F, Leduc M, Ludwig R (2019) Assessing natural variability in rcm signals: comparison of a multi model euro-cordex ensemble with a 50-member single model large ensemble. Climate Dynamics 53:1963--1979

\bibitem[{Ukkola et~al.(2018)Ukkola, Pitman, De~Kauwe, Abramowitz, Herger, Evans, and Decker}]{ukkola2018evaluating}
Ukkola A, Pitman A, De~Kauwe M, et~al (2018) Evaluating cmip5 model agreement for multiple drought metrics. Journal of Hydrometeorology 19(6):969--988

\bibitem[{Ukkola et~al.(2020)Ukkola, De~Kauwe, Roderick, Abramowitz, and Pitman}]{ukkola2020}
Ukkola AM, De~Kauwe MG, Roderick ML, et~al (2020) Robust future changes in meteorological drought in cmip6 projections despite uncertainty in precipitation. Geophysical Research Letters 47(11):e2020GL087820. \doi{https://doi.org/10.1029/2020GL087820}, \urlprefix\url{https://agupubs.onlinelibrary.wiley.com/doi/abs/10.1029/2020GL087820}, e2020GL087820 2020GL087820, {\href{https://arxiv.org/abs/https://agupubs.onlinelibrary.wiley.com/doi/pdf/10.1029/2020GL087820}{{https://agupubs.onlinelibrary.wiley.com/doi/pdf/10.1029/2020GL087820}}}

\bibitem[{Vicente-Serrano et~al.(2020)Vicente-Serrano, Quiring, Pe{\~n}a-Gallardo, Yuan, and Dom{\'\i}nguez-Castro}]{vicente2020review}
Vicente-Serrano SM, Quiring SM, Pe{\~n}a-Gallardo M, et~al (2020) A review of environmental droughts: Increased risk under global warming? Earth-Science Reviews 201:102953

\bibitem[{Vogel et~al.(2019)Vogel, Donat, Alexander, Meinshausen, Ray, Karoly, Meinshausen, and Frieler}]{vogel2019effects}
Vogel E, Donat MG, Alexander LV, et~al (2019) The effects of climate extremes on global agricultural yields. Environmental Research Letters 14(5):054010

\bibitem[{van~der Wiel et~al.(2023)van~der Wiel, Batelaan, and Wanders}]{van2023large}
van~der Wiel K, Batelaan TJ, Wanders N (2023) Large increases of multi-year droughts in north-western europe in a warmer climate. Climate Dynamics 60(5):1781--1800

\bibitem[{Zaveri et~al.(2023)Zaveri, Damania, and Engle}]{zaveri2023droughts}
Zaveri ED, Damania R, Engle NL (2023) Droughts and deficits-summary evidence of the global impact on economic growth. N/A

\end{thebibliography}
\nolinenumbers
\includepdf[pages=-]{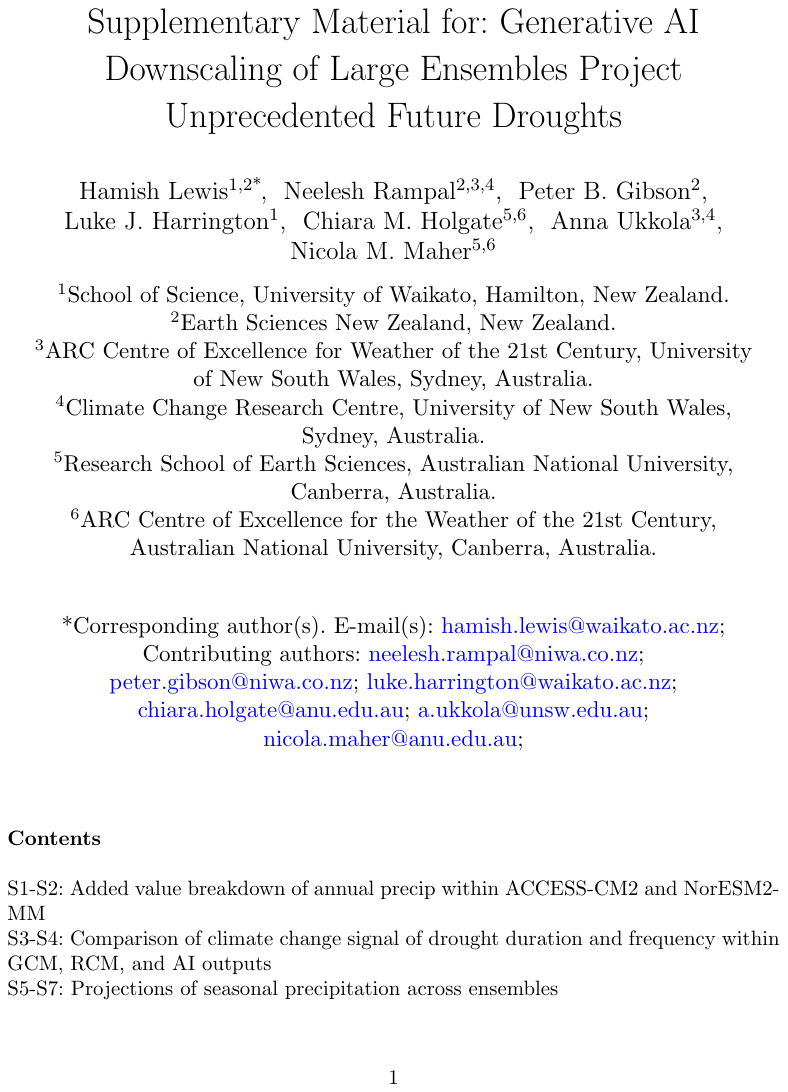}

\end{document}